\documentclass[a4paper,11pt]{article}
\usepackage[utf8]{inputenc}
\usepackage[T1]{fontenc}
\usepackage{comment}
\usepackage{tensor}
\usepackage{bbm}
\usepackage{float}
\usepackage{amssymb}
\usepackage{authblk}
\usepackage[dvipsnames]{xcolor}
\usepackage{graphicx}
\usepackage[new]{old-arrows}
\usepackage{jheppub}
\usepackage{bm}

\usepackage[multiple]{footmisc}
\usepackage{multicol}
\usepackage{multirow}
\usepackage{rotating}
\usepackage{array}
\usepackage{mathtools}
\usepackage{longtable}
\usepackage{booktabs}
\usepackage[mathscr]{euscript}
\usepackage{tikz}
\usetikzlibrary{backgrounds}
\usepackage{braket}
\usepackage{subfigure}
\usepackage{physics}
\usepackage{dsfont}
\usepackage{slashed}
\makeatletter
\gdef\@fpheader{\vphantom{Prepared for submission to JHEP}}
\makeatother

\usepackage{marginnote}

\newcommand{\RNum}[1]{\uppercase\expandafter{\romannumeral #1\relax}}

\newcommand{\betaobs}{\beta_{\rm obs}}
\tikzset{
	partial ellipse/.style args={#1:#2:#3}{
		insert path={+ (#1:#3) arc (#1:#2:#3)}
	}
}
\tikzset{snake it/.style={decorate, decoration=snake}}

\usetikzlibrary{calc,decorations.markings}
\usetikzlibrary{decorations.pathmorphing}
\usetikzlibrary{decorations.pathreplacing}
\usetikzlibrary{shapes,backgrounds}
\usetikzlibrary{fadings}
\usetikzlibrary{cd}
\usetikzlibrary{hobby}
\usetikzlibrary{decorations.pathreplacing}
\usetikzlibrary{knots}
\usetikzlibrary{3d}
\tikzfading
[
	name=fade out,
	inner color=transparent!0,
	outer color=transparent!100
]
\newif\ifdraft
\draftfalse

\newcommand{\be}{ \begin{equation}}
\newcommand{\ee}{\end{equation}}
\newcommand{\bi}{ \begin{itemize}}
\newcommand{\ei}{\end{itemize}}

\makeatletter
\newcommand*\bigcdot{\mathpalette\bigcdot@{.65}}
\newcommand*\bigcdot@[2]{\mathbin{\vcenter{\hbox{\scalebox{#2}{$\m@th#1\bullet$}}}}}
\makeatother

\title{\boldmath JT gravity on the worldline}
\author[*,\dagger]{Ben Freivogel,}
\author[*,\dagger]{Alessandro Fumagalli,}
\author[\dagger]{Diego M. Hofman}

\affiliation[*]{GRAPPA, University of Amsterdam, Science Park 904, 1090 GL Amsterdam, The Netherlands}
\affiliation[\dagger]{ITFA, University of Amsterdam, Science Park 904, 1090 GL Amsterdam, The Netherlands}

\emailAdd{freivogel@uva.nl, a.fumagalli@uva.nl, d.m.hofman@uva.nl}
\abstract{
Motivated by the problem of understanding the experience of an observer in dynamical quantum gravity, we study the effects of coupling a one-dimensional quantum mechanical system living on a bulk worldline to Euclidean AdS JT gravity. On the disk topology, where the worldline stretches between two boundary points, we derive exact expressions for the Euclidean propagator of the observer and for its correlation functions, and discuss their holographic interpretation. The main effect on the quantum mechanics is the fluctuation of the total Euclidean time for which the observer evolves, or its inverse temperature for closed Euclidean paths. This turns the standard quantum mechanical evolution operator into an average of those, weighted by a measure over Euclidean times which in the semiclassical limit is peaked around the geodesic distance between the boundary points. We characterize the fluctuations around this value, finding that they are small compared to the mean, but large compared to the effective Planck scale of the model. These fluctuations can be resolved by an observer with a finely spaced density of states. We also discuss the Lorentzian interpretation of these Euclidean calculations. Finally, we compute a contribution to the partition function of the observer coupled to gravity coming from the double trumpet. In this case the fluctuations of the effective temperature are large, reflecting the absence of a smooth semiclassical saddle point.
}

\arxivnumber{}

\begin{document}

\maketitle
\flushbottom
\begingroup\allowdisplaybreaks

\section{Introduction} \label{sec:introduction}

Describing the experience of an observer in the presence of dynamical gravity at the quantum level is a fundamental open problem. Although the notion of such an observer should be part of a complete theory of quantum gravity, we often work in settings where the latter can be described semiclassically. Our own situation in a universe that is entering an accelerated expansion epoch driven by a positive cosmological constant is one such case, since we can say that we live approximately on a geodesic. Even if the quantum gravitational effects that we could perceive are highly suppressed, understanding such effects in cosmological spacetimes, like ours, is key for comprehending the early universe at the microscopic level.
Hence, grasping the quantum gravity corrections to the experience of such an observer represents an important problem \cite{Goheer:2002vf, Banks:2003cg, Parikh:2004wh, Banks:2006rx}. Moreover, there are hints that it could be possible to formulate a holographic theory for cosmological spacetimes starting from such a worldline \cite{Anninos:2011af} (see also \cite{Anninos:2018svg, Anninos:2021ydw, Witten:2023xze,  Verlinde:2024znh, Verlinde:2024zrh,Narovlansky:2023lfz, Tietto:2025oxn, Goto:2026ipq, Blommaert:2026ofx} for recent progress in this direction), or from other features and boundaries that one can add to the static patch \cite{Anninos:2023epigravobs, Anninos:2024wpycosmoobs, Anninos:2024xhcadsbdy, Banihashemi:2026mje, Philcox:2025faf, Silverstein:2024xnr, Batra:2024kjl,Coleman:2021nor}. In both cases, one has to deal with the interaction of gravity with the dual theory residing on the worldline or the boundary. As with AdS/CFT \cite{Maldacena:1997re, Witten:1998qj}, refining these ideas would help us sharpen some of the techniques we employ, for instance, to study initial conditions through the wavefunction of the universe \cite{Hartle:1983ai} (see \cite{Maldacena:2019cbz, Chen:2020tes, Fumagalli:2024msi, Ivo:2024ill, Anninos:2024iwf, Anninos:2025fer, Abdalla:2026mxn, Harlow:2026hky, Zhao:2026mpl} for recent discussion and applications), which lack a nonperturbative definition.

Besides the cosmological application, studying quantum gravitational effects on a bulk observer is obviously an interesting subject per se, the case of a black hole environment being a prime example. Here, questions on horizon physics and firewalls \cite{Mathur:2009hf, Almheiri:2012rt} have been understood to a certain degree \cite{Penington:2019npb, Almheiri:2019psf, Almheiri:2019qdq, Penington:2019kki, Blommaert:2019hjr}. However, the description of an infalling observer inside a black hole is still rather confusing (see \cite{Jafferis:2020ora, deBoer:2022zps, Leutheusser:2021frk, Leutheusser:2021qhd, Mertens:2025rpa,Franken:2026bff} for some direct approaches).

Many recent works have explained in detail how including observers is important for discussing gravitational physics in situations where there is no weakly gravitating region to dress observables, see for example \cite{Leutheusser:2021frk, Witten:2023background, Chandrasekaran:2022cip, Mirbabayi:2023vgl, Kolchmeyer:2024fly, Abdalla:2025gzn, Akers:2025ahe, Speranza:2025joj, Yang:2025lme}. Instead, we are more interested in the effect of gravitational interactions on the observer. Other related works can be found in \cite{Nitti:2024iyj, Banerjee:2024ioz}.

In particular, we take a step toward describing the effects of quantum gravity on a quantum mechanical observer living in the bulk of spacetime, where gravity is not assumed to be weak. We work in a two-dimensional model, namely Jackiw-Teitelboim gravity \cite{Jackiw:1984je, Teitelboim:1983ux, Almheiri:2014cka, Engelsoy:2016xyb,Maldacena:2016hyu}. This theory has no local gravitational degrees of freedom. However, there are two types of quantum gravitational effects that make it interesting. The first is the presence of a dynamical fluctuating boundary (the "boundary graviton"); the second is given by topology change, which is under control in this theory, and significantly affects many observables \cite{Saad:2018bqo, saad2019jt, Saad:2019pqd, Penington:2019npb, Almheiri:2019psf, Almheiri:2020cfm,Penington:2019kki}. As a first step, we will focus on the former type of effect and thus work at fixed topology. We consider a worldline characterized by a mass $m$ (one can think of it as a laboratory), on top of which we place a one-dimensional quantum mechanical degree of freedom (which can be thought of as an experiment performed inside the lab), which is the observer; we couple this system to the boundary mode of JT gravity. We first work in Euclidean signature on the disk topology, finding that the coupling to gravity induces fluctuations of the total Euclidean time for which the observer lives
similar to the one that one would have when turning on one-dimensional quantum gravity on a worldline \cite{Anninos:2021ydw}. The evolution of the quantum system is given by an integral over the standard quantum mechanical evolution operator $U_{QM}(\betaobs)$ for different Euclidean times $\betaobs$\footnote{We use the notation $\betaobs$ for  Euclidean time even when the paths are not periodic.}
\begin{align}\label{eq:Uaverage}
	U_{QG}=\int d\betaobs\, \mu(\betaobs) \,U_{QM}(\betaobs)=\int d\betaobs\, \mu(\betaobs) e^{-\betaobs H_{QM}}.
\end{align}
This average enters quantities such as the partition function, the Euclidean propagator and correlation functions. The measure for this integral is induced by the gravitational theory in ambient space and admits a closed exact expression. We show that in situations where there is a semiclassical saddle point, the measure $\mu(\betaobs)$ is peaked around a particular value that is set by the equations of motion, thus recovering the standard evolution of the quantum system. We characterize fluctuations around this semiclassical limit, and we find that they come from two sources: transverse fluctuations of the worldline, which would exist even without gravity, and a genuine quantum gravity contribution.
On the disk topology, where the worldline extends between two points on the boundary, the first term turns out to be the dominant one, enhanced with respect to the gravitational one by a logarithmic power of the IR cutoff of JT gravity. However, we show that an observer with a fine-grained density of states can still experience such effects. Interestingly, the size of the quantum gravity contribution when the observer backreacts weakly scales with a combination of the effective Planck length in JT gravity $\beta/\phi_r$ and the AdS radius
\begin{align}
    \delta \beta_{\rm{obs}, QG}^2= \frac{\beta\, r_{\rm{AdS}}}{\phi_r \pi^2}.
\end{align}
This is larger than the effective Planck length squared, reminiscent of what has been observed in \cite{Marolf:2003bb, Verlinde:2019ade,Verlinde:2019xfb, Verlinde:2022hhs, Freivogel:2026ofo, Freivogel:2026bsx}. We also describe the holographic dual of our bulk calculation, and discuss holographic renormalization. This allows us to comment on the Lorentzian interpretation of the results: the propagator of the observer between two points at the boundary of thermal Lorentzian AdS$_2$ is computed through an analytic continuation of equation
\eqref{eq:Uaverage}
\begin{align}
  U_{QG}=\int d t\, \mu(t) \,U_{QM}(t)=\int dt\, \mu(t) \,e^{-i t H_{QM}}.
\end{align}
Notice that, if we interpret $U_{QM}(t)$ as unitary quantum evolution for our quantum mechanical worldline degree of freedom, the average above does not yield unitary worldline evolution.
This is familiar in the worldline formalism for QFT amplitudes where Schwinger parameterization of propagators can be used to connect this calculation to the physics of a unitary S-matrix through the LSZ formalism. Here, the situation is further complicated by the fact that boundary graviton fluctuations can be interpreted as a sort of average in the dual theory, as is realized in the SYK model where we must average over couplings \cite{Maldacena:2016hyu, saad2019jt}. In such a scenario, we do not even expect the boundary theory to allow for unitary evolution. On top of all this, in AdS/CFT in the semiclassical limit (i.e. large operator dimensions) worldline saddles are dominated by complex geodesics in Lorentzian time; that become real geodesics in the Euclidean setup. This relates directly to the confining AdS potential for massive particles and connects to holographic renormalization. This is certainly true in our case where the analytically continued measure $\mu(t)$ has a saddle point that is complex.

The upshot is that the composition of all these effects casts doubt on whether one should expect the observer's experience to be unitary in Quantum Qravity. The difficulty of defining bulk observables in general is the major obstruction to trying to settle this question in general setups. The particular scenario we study in this work is simple enough that we hope we can shed some light on this issue: first by studying boundary anchored observables (including bulk worldline operator insertions) on the disk topology and then by considering closed geodesics in double-trumpet geometries. In this last case, fluctuations of Euclidean time are interpreted as fluctuations of the temperature felt by the observer, and we find that the induced inverse temperature is not sharply peaked in either of the limits we study. This occurs because, in pure gravity with the worldline and the observer, there is no smooth classical saddle controlling the double-trumpet path integral, and thus observables receive contributions from many off-shell geometries.

We conclude with a discussion of the results and several future directions.

\section{General framework} \label{sec:generalframework}
In this section, we introduce our setup and clarify notation. We consider a gravitational theory coupled to a quantum mechanical system living on a bulk worldline. We take the gravitational theory to be Jackiw-Teitelboim gravity \cite{Jackiw:1984je,Teitelboim:1983ux,Almheiri:2014cka,Engelsoy:2016xyb,Maldacena:2016upp}, a two-dimensional dilaton gravity theory that describes the near-horizon limit of four-dimensional near-extremal black holes. The Euclidean action reads
\begin{align}
   I_{JT}= - \frac{\phi_0}{2}\left[\int R+ 2\int_\partial K\right]-\frac{1}{2} \left[\int \phi\left(R+\frac{2}{r_{\rm{AdS}}^2}\right)+2 \int_\partial \phi_b K\right]
\end{align}
We will set $r_{\rm{AdS}}=1$ in this work, unless stated otherwise. When we do so, the bulk coordinates become dimensionless. The first term is purely topological, and we will not consider it because we work at fixed topology. The dilaton $\phi$ plays the role of the two-dimensional (dimensionless) inverse gravitational constant. It appears as a nondynamical Lagrange multiplier; we can integrate it out and be left with a geometry that is locally EAdS$_2$. After cutting off the geometry close to the boundary, the remaining gravitational dynamics governs reparametrizations of the boundary curve. The cutoff geometry is specified by the boundary value of the dilaton
$\phi|_\partial =\phi_b$ and the length of the boundary curve\footnote{More general classes of boundary conditions in JT gravity were used in various contexts, from spectral form factor computations \cite{Saad:2018bqo} to toy models of cosmology \cite{Fumagalli:2024msi} constructed from the de Sitter version of JT gravity \cite{Maldacena:2019cbz}. A detailed classification is described in \cite{Goel:2020yxl}.}  $\beta_b=\int_\partial \sqrt{h}$.

In the limit where this curve is very close to the AdS boundary, the dynamics of the curve is described by the Schwarzian action \cite{Maldacena:2016upp}, and the theory has a nearly CFT boundary description \cite{Maldacena:2016hyu}. This limit can be reached by taking both $\phi_b$ and $\beta_b$ large, keeping their ratio fixed. It is standard to define renormalized values for these objects, $\phi_b= \phi_r/\epsilon$, $\beta_b= \beta/\epsilon$, and take $\epsilon\ll 1$ (and therefore $\phi_r,\beta$ must not scale with $\epsilon$). In these variables, the dynamics of the boundary curve is described by the Schwarzian action \cite{Maldacena:2016upp}. If we consider the disk topology and choose Poincaré coordinates
\begin{align}
	ds^2=\frac{dt^2+dz^2}{z^2}\,,
\end{align}
the path integral over the JT degrees of freedom runs over the different boundary curves $x_b^\mu(u)=(t(u),z(u))$ constrained by a condition on the boundary metric, which by a choice of the $u$ coordinate reads
\begin{align}
	g_{uu}du^2=\frac{du^2}{\epsilon^2}=\frac{t'(u)^2+z'(u)^2}{z(u)^2}.
\end{align}
Using this condition, one can rewrite $x_b^\mu(u)=(t(u),z(u))\simeq (t(u),\epsilon t'(u))$, where $u\in (0,\beta)$ is a coordinate that runs on the boundary, and the action becomes
\begin{align}\label{eq:JTaction}
	I_{JT}= \text{topological}-\frac{\phi_r \beta}{\epsilon^2}- \phi_r\int_0^\beta du \{t(u),u\} +\order{\epsilon^2}
\end{align}
where $\{t,u\}=\frac{t'''}{t'}-\frac{3}{2}\frac{t''^2}{t'^2}$ is the Schwarzian derivative. This action has an $SL(2,\mathbb R)$ gauge group acting on the coordinate $t(u)$, which corresponds to AdS transformations that leave the cutoff geometry invariant. It is customary to renormalize away the constant $\phi_r \beta/\epsilon^2$ term by adding a local volume term in the action; we will drop such a term from now on. If one looks only at the Schwarzian action, all quantities depend only on the ratio $\beta_b/\phi_b=\beta/\phi_r$. The gravitational semiclassical regime is reached when one takes $\phi_r$ large, corresponding to the limit where the length scale $\beta/\phi_r $ is sent to zero (this can be thought of as an effective dimensionful Planck length of the theory). Finite cutoff effects in \eqref{eq:JTaction} are suppressed by powers of $\epsilon^2/\beta^2$, so we must ensure that $\beta$ does not scale with the cutoff in order to trust the Schwarzian action. We will see later that, when including the observer, one needs additional constraints: first, the effective Planck length must satisfy $\epsilon \phi_r/\beta \ll 1$; second, the energy of the worldline plus the observer must satisfy $\epsilon E \ll 1$\footnote{Similar restrictions appear when computing boundary correlators in JT. Within the Schwarzian theory, one must consider matter operators whose conformal dimensions do not scale with the cutoff.}. One can define a new field $t(u)=\tan(f(u) \frac{\pi}{\beta})$ that winds around the $u$ thermal circle: $f(u) \sim f(u)+\beta$. A classical solution is $f(u)=u$, which winds once around the thermal circle in $u$ and describes a black hole of inverse temperature $\beta$. Since everything will depend on the ratio $\beta/\phi_r$, for simplicity of presentation we will rescale the inverse temperature $\beta/\phi_r \to \beta$ and rescale the $u$ coordinate in the same way so that $u \in (0, \beta)$.

We now want to add the degrees of freedom describing the worldline $X_w^\mu(\tau)$ (where $\tau$ is an affine parameter running on it), and the observer, which we take to be a one-dimensional quantum mechanical degree of freedom living on the worldline, encoded by the variable $q(\tau)$. The setup is depicted in figure \ref{fig:EAdS_disk}.
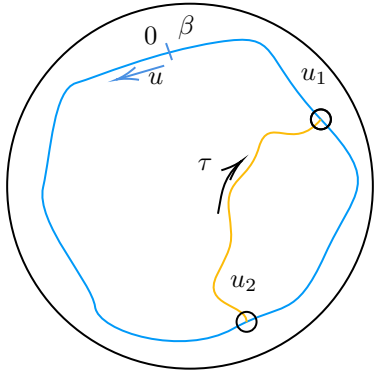
\begin{figure}
    \centering

\tikzset{every picture/.style={line width=0.75pt}} 

\begin{tikzpicture}[x=0.75pt,y=0.75pt,yscale=-1,xscale=1]

\draw   (61,151.5) .. controls (61,100.97) and (101.97,60) .. (152.5,60) .. controls (203.03,60) and (244,100.97) .. (244,151.5) .. controls (244,202.03) and (203.03,243) .. (152.5,243) .. controls (101.97,243) and (61,202.03) .. (61,151.5) -- cycle ;
\draw  [color={rgb, 255:red, 1; green, 153; blue, 248 }  ,draw opacity=1 ] (166.5,79.25) .. controls (198.5,74.25) and (187,85) .. (218,118) .. controls (249,151) and (235,153) .. (220,186) .. controls (205,219) and (199,209) .. (174,223) .. controls (149,237) and (106.75,224.75) .. (105,212) .. controls (103.25,199.25) and (75,170) .. (79,152) .. controls (83,134) and (84.13,103.13) .. (95.5,99.75) .. controls (106.88,96.38) and (134.5,84.25) .. (166.5,79.25) -- cycle ;
\draw [color={rgb, 255:red, 252; green, 187; blue, 9 }  ,draw opacity=1 ]   (218,118) .. controls (204,135) and (192,115) .. (187,132) .. controls (182,149) and (170,147) .. (172,164) .. controls (174,181) and (174,180) .. (166,196) .. controls (158,212) and (181.2,210.5) .. (180.8,219.7) ;
\draw [color={rgb, 255:red, 74; green, 144; blue, 226 }  ,draw opacity=1 ]   (139.58,91.08) -- (117.5,97.52) ;
\draw [shift={(115.58,98.08)}, rotate = 343.74] [color={rgb, 255:red, 74; green, 144; blue, 226 }  ,draw opacity=1 ][line width=0.75]    (10.93,-3.29) .. controls (6.95,-1.4) and (3.31,-0.3) .. (0,0) .. controls (3.31,0.3) and (6.95,1.4) .. (10.93,3.29)   ;
\draw    (166.5,165.33) .. controls (168.4,155.2) and (168.5,148.38) .. (177.07,140) ;
\draw [shift={(178.5,138.67)}, rotate = 138.01] [color={rgb, 255:red, 0; green, 0; blue, 0 }  ][line width=0.75]    (10.93,-3.29) .. controls (6.95,-1.4) and (3.31,-0.3) .. (0,0) .. controls (3.31,0.3) and (6.95,1.4) .. (10.93,3.29)   ;
\draw [color={rgb, 255:red, 74; green, 144; blue, 226 }  ,draw opacity=1 ]   (143.5,88.25) -- (140.5,79.75) ;

\draw (155,136) node [anchor=north west][inner sep=0.75pt]  [font=\small] [align=left] {$\displaystyle \tau $};
\draw (130,93) node [anchor=north west][inner sep=0.75pt]  [font=\small] [align=left] {$\displaystyle u$};
\draw (206.4,91.2) node [anchor=north west][inner sep=0.75pt]  [font=\small]  {$u_{1}$};
\draw (171.1,194.9) node [anchor=north west][inner sep=0.75pt]  [font=\small]  {$u_{2}$};
\draw (128,69.9) node [anchor=north west][inner sep=0.75pt]  [font=\small]  {$0$};
\draw (145,64.4) node [anchor=north west][inner sep=0.75pt]  [font=\small]  {$\beta $};

\draw   (218, 118) circle [x radius= 5, y radius= 5]   ;
\draw   (218, 118) circle [x radius= 5, y radius= 5]   ;
\draw   (180.8, 219.52) circle [x radius= 5, y radius= 5]   ;
\end{tikzpicture}

    \caption{Euclidean AdS with a worldline (yellow). The worldline is attached to the wiggly boundary of cutoff EAdS$_2$, which has fixed length $\beta_b=\beta/\epsilon$. Another physical parameter that will enter computable quantities is the difference $u_1-u_2$.}
    \label{fig:EAdS_disk}
\end{figure}
We take their action to be
\begin{align}\label{eq:observeractiongeneral}
	I_{obs}=I_{worldline}+I_{QM}= m\int_0^1 d \tau \,e_\tau\left(e_\tau^{-2}\, \frac{\Dot{X}_w^2}{2}+\frac{1}{2}\right)+ \int_0^1 d \tau e_\tau \mathcal L_{QM} [q(\tau),e_\tau]\,,
\end{align}
where we make use of an auxiliary worldline einbein $e_\tau$, denote $\Dot{X}_w^2=g_{\mu\nu}\Dot{X}_w^\mu\Dot{X}_w^\nu$, and use dots for derivatives with respect to the worldline time $\tau$. We will keep the quantum system generic for most of this work, assuming that is described by the Lagrangian
\begin{align}
	\mathcal L_{QM}[q(\tau),e_\tau]= \frac{\dot q^2}{2 e_\tau}+ e_\tau V(q)\,.
\end{align}
Since the bulk metric is fixed to be EAdS$_2$, the coupling to the gravitational degrees of freedom is only through the boundary curve, as described in figure \ref{fig:EAdS_disk}. It is implemented by means of the boundary conditions of the worldline endpoints:
\begin{align}\label{eq:Xboundarycond}
	&X_w^\mu(\tau=0)=x_b^\mu(u_2)=(t(u_2),z(u_2))\,,\quad 	X_w^\mu(\tau=1)=x_b^\mu(u_1)=(t(u_1),z(u_1))\,.
\end{align}
We can also choose the $u$ coordinates in such a way that $u_2=0$. One must also set boundary conditions on the quantum mechanical degree of freedom at the endpoints of the worldline; we consider Dirichlet conditions:
\begin{align}\label{eq:qboundarycond}
	q(\tau=0)=q_i\,,\quad q(\tau=1)=q_f\,,
\end{align}
where the subscripts $i,f$ stand for initial and final. These quantities will also be arguments of any computable quantity of interest. They define the initial and final states of the quantum system. It is also straightforward to generalize this construction to include generic initial and final quantum states $\ket{\psi_i},\ket{\psi_f}$, as we will explain later. The results we find throughout this work hold for any quantum system coupled to gravity in the way we described; however, for concreteness we can think of working with a simple harmonic oscillator\footnote{Another possible minimal choice of quantum system is a clock degree of freedom introduced and studied in various works; see, for example, \cite{Chandrasekaran:2022cip, Witten:2023xze, Mirbabayi:2023vgl, Kolchmeyer:2024fly, Yang:2025lme}.}. The full worldline and observer action is therefore
\begin{align}\label{eq:observeraction}
	I_{obs}&=I_{worldline}+I_{observer}= \nonumber\\
	&=m\int_0^1 d \tau \,e_\tau\left(e_\tau^{-2}\, \frac{\Dot{X}_w^2}{2}+\frac{1}{2}\right)+ \int_0^1 d \tau \,e_\tau\left(e_\tau^{-2}\, \frac{ \Dot{q}^2}{2}+ \frac{ \omega^2 {q}^2}{2}\right)\,.
\end{align}
When we do not set $r_{\rm{AdS}}=1$, the units we are using on the worldline are
\begin{align}\label{eq:worldlineunits}
	[1/m]=[X_w]=[q^2]=[e_\tau]=[\omega]^{-1}=\text{length}\,,\quad [\tau]=(\text{length})^0
\end{align}
Notice that $\omega$ is the only new scale introduced by the quantum system, since any other scale can be removed by redefining $q$. If we set the AdS radius to be unit valued, then all these quantities become dimensionless. Reintroducing units is straightforward: we just insert powers of $r_{\rm{AdS}}$ to match \eqref{eq:worldlineunits}. It is worth mentioning that this is not the only gauge invariant action one can write. We will discuss other options and their relation to this one in appendix \ref{app:differentactions}.
\section{Observer and quantum gravity on the disk}
\subsection{The Euclidean propagator: an exact formula}\label{sec:diskpropagator}
In the Euclidean JT gravity setup described in figure \ref{fig:EAdS_disk}, the worldline on top of which the observer lives extends between two points on the fluctuating boundary. From the point of view of the observer, we are describing the evolution between two states $\ket{q_i},\ket{q_f}$ prepared at the start and end of the worldline trajectory, with a Euclidean time that fluctuates because of the dynamics of the boundary graviton and the wiggly worldline. With this choice of initial and final states, the quantity we are computing is the Euclidean quantum mechanical propagator between these two points, dressed with the gravitational mode of JT gravity. In Euclidean quantum mechanics this evolution is encoded by a Hermitian operator (corresponding to unitary one in Lorentzian signature) $\mel{q_f}{U_{QM}(\betaobs)}{q_i}$ that depends on some fixed Euclidean time $\betaobs$. In our gravitational setup, the propagator is instead computed by the following path integral
\begin{align}\label{eq:propagatorfirst}
	&\bra{q_f} U_{QG}(\beta,u_1,m)\ket{q_i}=\\
	&=\int \mathcal De_\tau \int \mathcal Dx_b^\mu(u) \,e^{-I_{JT}[x_b^\mu(u)]}\int_{X_w^\mu(\tau=0) = x_b^\mu(u=0)}^{X_w^\mu(\tau=1) =   x_b^\mu(u=u_1)}\mathcal{D}{X_w(\tau)} \int_{q_i,\tau=0}^{q_f,\tau=1} \mathcal D q  e^{-I_{obs}[X_w^\mu(\tau),q(\tau),e_\tau]}\,.\nonumber
\end{align}
where we have made explicit the boundary conditions of the $X_w$ and $q$ path integrals; the coordinate ranges are $\tau \in (0,1)$ and $u \in (0,\beta)$. Notice that this quantum gravity evolution operator $U_{QG}$ depends on various parameters that are external to the observer. In particular, none of these parameters can be straightforwardly identified with the Euclidean time on the worldline. The path integral over the JT degrees of freedom runs over the different boundary curves $x_b^\mu(u)=(t(u),z(u))$, and, as we discussed in \ref{sec:generalframework}, can be parameterized by a path integration over a single field $t(u)$. We prefer to use the former notation for reasons that will be clearer later. Another important feature of the JT path integral is that we must divide by the volume of the gauge group $SL(2,\mathbb R)$. We leave this implicit in $\mathcal D x_b(u)$, but it will be important in a later step. One can also smear the states $\ket{q_{i,f}}$ with wavefunctions in position space to generalize this formula to generic initial and final quantum states
\begin{align}
	\bra{\psi_f} U_{QG}(\beta,u_1,m)\ket{\psi_i}=\int dq_i dq_f\, \psi_i(q_i) \psi_f^*(q_f) \bra{q_f} U_{QG}(\beta,u_1,m)\ket{q_i}.
\end{align}

It turns out that in this simple model of quantum gravity these path integrals can all be performed exactly, giving expressions written as ordinary integrals. We will now show how to do so. The path integral over the einbein becomes trivial once we gauge fix it to be $e_\tau=\betaobs$, where $\betaobs=\int_0^1 d\tau\,e_\tau$ is the proper Euclidean time on the worldline. The Faddeev-Popov determinant is a numerical constant (as we show in appendix \ref{app:einbeinggaugefixing}), and the einbein path integral reduces to an integral over the zero mode $\betaobs$ (we will drop numerical prefactors from now on)
\begin{align}\label{eq:intermediateb}
	&\bra{\psi_f} U_{QG}(\beta,u_1,m)\ket{\psi_i}= \nonumber \\
	&=\int_0^\infty d\betaobs\, e^{-\frac{\betaobs m}{2}} \int \mathcal Dx_b^\mu(u) \,e^{-I_{JT}[x_b^\mu(u)]}\nonumber\\
	&\quad\times
	\int_{X_w^\mu(0) = x_b^\mu(0)}^{X_w^\mu(\betaobs) = x_b^\mu(u_1)}
	\mathcal{D}{X_w(\tau)} e^{-\int_0^{\betaobs} d\tau \frac{m\dot X_w^2}{2}}
	\mel{\psi_f}{e^{-\betaobs H}}{\psi_i}\,,
\end{align}
where we have rescaled the worldline coordinate so that it runs over an interval of length $\betaobs$. The path integral over the particle worldline $X_w^\mu$ is nothing but the untraced heat kernel in AdS between the two endpoints $x_b^\mu(u=0)$ and $x_b^\mu(u=u_1)$, and can be reduced to an integral \cite{Camporesi:1990wm, Bergamin:2015vxa}. Renaming these two boundary points $x_i^\mu,x_f^\mu$, we are left with
\begin{align}
	&\bra{\psi_f} U_{QG}(\beta,u_1,m)\ket{\psi_i}= \nonumber \\
	&=\int_0^\infty \frac{d\betaobs}{\betaobs^{3/2}} e^{-\frac{\betaobs m}{2}-\frac{\betaobs}{8m}}   \int \mathcal Dx_b^\mu(u) \,e^{-I_{JT}[x_b^\mu(u)]}\int_{d(x_i, x_f)}^{\infty} \frac{dy \, y \, e^{-\frac{m y^2}{2\betaobs}}}{\sqrt{\cosh y - \cosh d(x_i, x_f)}} \times \nonumber\\
	&\times \mel{\psi_f}{e^{-\betaobs H}}{\psi_i},
\end{align}
where $d(x_i,x_f)$ is the geodesic distance between the two points. In Poincaré coordinates $x^\mu=(t,z)$ this is
\begin{align}
	\cosh d(x_i,x_f)=1+\sigma^2=1+\frac{(t_{i}-t_f)^2+(z_{i}-z_f)^2}{2z_iz_f} \simeq \frac{(t_{i}-t_f)^2}{2\epsilon^2}\,,
\end{align}
where $\sigma$ is sometimes called the chordal distance between the two points, and in the last step we expanded for small cutoff. Here $y$ is the heat-kernel proper-length variable: $y=d(x_i,x_f)$ is the geodesic saddle, while $y>d(x_i,x_f)$ parameterizes off-geodesic worldline paths. We will sometimes call these QFT or Brownian fluctuations, given the analogous diffusion scale $\betaobs/m$.

We must now do the path integral over the JT degrees of freedom. Before doing so, let us note the following. If we did not have the quantum system, the quantity that we are computing would be the path integral of a particle of mass $m$ in EAdS, dressed with the quantum gravity mode of JT. In this case, the integral over the worldline degrees of freedom gives the propagator of a massive scalar field in AdS between the two bulk points $x_f^\mu=x_b^\mu(u_1),x_i^\mu=x_b^\mu(0)$. Thus this is nothing but the calculation of the two point function of a scalar of mass $m$ on the boundary of the cutoff geometry, which in the AdS/CFT dictionary becomes a conformal operator of dimension $\Delta=\frac{1}{2}+\sqrt{\frac{1}{4}+m^2}$ dressed with the boundary gravitational mode. The result of the path integral in this case is known exactly and can be written as ordinary integrals \cite{Yang:2018gdb, Mertens:2017mtv}. The results from these works can be used to compute the JT path integral even in our slightly more complicated setup. To do so, we split the path integral over the wiggly boundary in two at the points where the worldline is anchored to the boundary (call them $x_i^\mu, x_f^\mu$):
\begin{align}
	&\bra{\psi_f} U_{QG}(\beta,u_1,m)\ket{\psi_i}= \nonumber \\
	&=\int_0^\infty \frac{d\betaobs}{\betaobs^{3/2}} e^{-\frac{\betaobs m}{2}-\frac{\betaobs}{8m}} \int d^2x_i \int d^2x_f \int_{x_i}^{x_f} \mathcal{D}x_b^\mu(u) e^{-I_{JT}[x_b^\mu(u)]} \int_{x_f}^{x_i} \mathcal{D}x_b^\mu(u) e^{-I_{JT}[x_b^\mu(u)]}\times\\
	&\times\int_{d(x_i, x_f)}^{\infty} \frac{dy \, y \, e^{-\frac{my^2}{2\betaobs}}}{\sqrt{\cosh y - \cosh d(x_i, x_f)}} \mel{\psi_f}{e^{-\betaobs H}}{\psi_i}.
\end{align}
Here, $\int d^2x_i$ denotes integrals over the full AdS space with the appropriate volume form. The path integrals of the boundary mode between two points have been computed in \cite{Yang:2018gdb} and, up to phases that cancel in our calculation, yield Wheeler-de Witt wavefunctions. Moreover, we can exploit the $SL(2,\mathbb R)$ gauge invariance to get rid of three of these integrals and fix $x_i^\mu=(0,\epsilon)$ and $x_f^\mu=(\ell,\epsilon)$, where $\epsilon$ is the IR cutoff and $t_f=\ell$ is the only remaining integration variable. We arrive at the following formula
\begin{align}\label{eq:QGpropagatorfinal}
	&\bra{\psi_f} U_{QG}(\beta,u_1,m)\ket{\psi_i}=\int_0^\infty \frac{d\betaobs}{\betaobs^{3/2}} e^{-\frac{\betaobs m}{2}-\frac{\betaobs}{8m}} \int_0^\infty d\ell \,\ell\,\varphi_{u_1}(\ell)\varphi_{\beta-u_1}(\ell)\times\\
	&\times\int_{d(\ell)}^{\infty} \frac{dy \, y \, e^{-\frac{my^2}{2\betaobs}}}{\sqrt{\cosh y - \cosh d(\ell)}} \mel{\psi_f}{e^{-\betaobs H}}{\psi_i}\,,
\end{align}
where $\cosh d(\ell)=1+\frac{\ell^2 }{2 \epsilon^2}$, and $\varphi(\ell)$ are the Wheeler-de Witt wavefunctions of JT gravity in the $\ell$ basis \cite{Yang:2018gdb}\footnote{The wavefunctions in \cite{Yang:2018gdb} have an additional $\epsilon$ prefactor that cancels with the AdS$_2$ volume factors; we avoid writing it directly in our formula.}
\begin{align}\label{eq:wdws}
	\varphi_u(\ell)=\frac{2}{\pi^2\ell}\int_0^\infty ds\, s\, \sinh(2 \pi s)e^{-\frac{s^2\,u}{2}}K_{2is}\left(\frac{4}{\ell}\right)\,.
\end{align}
These wavefunctions are obtained after taking the cutoff to zero while keeping the renormalized chordal distance $\ell$ fixed. Their use in \eqref{eq:QGpropagatorfinal} is therefore self consistent only when the dominant configurations remain far from the cutoff scale\footnote{In these integrals, this is guaranteed by the small $\ell$ behaviour of the Bessel function $K_{2is}(4/\ell) \approx e^{4/\ell}$.}; at the saddle point this requires $\ell^*/\epsilon\gg1$, giving the further constraints anticipated in the introduction and toroughly discussed in section \ref{sec:saddlepoint}. It is possible to use an integral representation of the Bessel function to rewrite these wavefunctions in another representation, which is more suitable for taking saddle points:
\begin{align}\label{eq:wdwtheta}
	\varphi_u(\ell)=\frac{\sqrt{2}}{4\pi^{3/2}u^{3/2}}\frac{1}{\ell}\int_{c+i\mathbb{R}} d\theta \, \theta\, e^{ \frac{\theta^2}{2u}+\frac{4}{\ell} \cos \theta/2}\,,
\end{align}
where the defining contour has $c=\pi$.

We have derived an exact formula for the Euclidean propagator of the observer.
Let us take the result \eqref{eq:QGpropagatorfinal} and write it in the following way:
\begin{align}\label{eq:propagatormuintegral}
	\bra{\psi_f} U_{QG}(\beta,u_1,m)\ket{\psi_i}=\int d\betaobs \mu_{\beta,u_1,m}(\betaobs)\mel{\psi_f}{e^{-\betaobs H}}{\psi_i}\,.
\end{align}
It is clear from this equation what the effect of coupling quantum mechanics to quantum gravity in ambient space and to the worldline wiggles is: it introduces an integral over the total elapsed Euclidean time, with a measure $\mu_{\beta,u_1,m}(\betaobs)$ that depends on the gravitational degrees of freedom and the mass of the worldline. Notice that the quantity we computed still depends on the IR cutoff $\epsilon$. The dependence of the measure on this cutoff is physical since it affects the proper times $\betaobs$ for which the observer is evolved. We will therefore keep this quantity in our expressions, and discuss the relation to holographic renormalization in the next section.
 \subsection{Holographic renormalization and dual description}\label{sec:renormalization}
In this section we reconsider the propagator computed above and derive it in another way. This will allow us to discuss holographic renormalization and the holographic interpretation of the previous bulk calculation.

Let us expand the quantum mechanical states in the energy basis:
\begin{align}
	\ket{\psi_i}= \sum_n \psi_i(E_n) \ket{E_n}.
\end{align}
Here, $\ket{E_n}$ is an energy eigenstate and $\psi_i(E_n)=\braket{E_n}{\psi_i}$ is the overlap with the state $\ket{\psi_i}$. Let us go back to the propagator written as \eqref{eq:intermediateb}; after inserting this expansion and using the fact that the Hamiltonian is diagonal in this basis, we obtain
\begin{align}
	\bra{\psi_f}U_{QG}\ket{\psi_i}=&\sum_n\int_0^\infty d\betaobs e^{-\frac{\betaobs m}{2}} \int \mathcal Dx_b^\mu(u) \,e^{-I_{JT}[x_b^\mu(u)]}\times\\
 &\times \int_{X_w^\mu(\tau=0) = x_b^\mu(u=0)}^{X_w^\mu(\tau=\betaobs) =   x_b^\mu(u=u_1)}\mathcal{D}{X_w(\tau)} e^{-m\int_0^{\betaobs} d\tau \frac{\dot X_w^2}{2}}\times \psi_f(E_n)^* \psi_i(E_n) e^{-\betaobs E_n}\,.
\end{align}
Now the $\betaobs$ and $X_w$ path integrals can be computed exactly, and they yield a two point function of scalar fields that are dual to boundary CFT operators of dimension $\Delta_n=\frac{1}{2}+\sqrt{\frac{1}{4}+m^2+2 mE_n}$ anchored at the wiggly boundary\footnote{The nonlinear relation between the new mass squared and the energy is due to the different notions of worldline energy and target space energy with our coupling of the observer through the einbein. For small energies instead the energy simply shifts the mass $m \to m+E_n$. More details on this can be found in appendix \ref{app:differentactions}.}. We therefore get
\begin{align}\label{eq:qftsum}
	\bra{\psi_f}U_{QG}\ket{\psi_i}=&\sum_n \psi_f(E_n)^* \psi_i(E_n)\int \mathcal Dx_b^\mu(u) \,e^{-I_{JT}[x_b^\mu(u)]}  \expval{\mathcal{O}^{\Delta_n}(x^\mu(0))\mathcal{O}^{\Delta_n}(x^\mu(u_1))}_{\rm QFT}\,,
\end{align}
where $\expval{\mathcal{O}^{\Delta_n}(x^\mu(0))\mathcal{O}^{\Delta_n}(x^\mu(u_1))}_{\rm QFT}$ denotes the bulk to bulk two point function of the bulk fields corresponding to the boundary operators $\mathcal O^{\Delta_n}$, anchored to the wiggly boundary. For each of these correlators, it is known how to compute the effect of the fluctuations of the boundary mode as an integral over WdW wavefunctions \cite{Yang:2018gdb, Mertens:2017mtv}. Calling the result $\expval{\mathcal{O}^{\Delta_n}(0)\mathcal{O}^{\Delta_n}(u_1)}_{QG}$, we can write the final expression as
\begin{align}\label{eq:observerpropagatorCFT}
	\bra{\psi_f}U_{QG}\ket{\psi_i}=&\sum_n \epsilon^{2 \Delta_n} \psi_f(E_n)^* \psi_i(E_n)\expval{\mathcal{O}^{\Delta_n}(0)\mathcal{O}^{\Delta_n}(u_1)}_{QG}\,,
\end{align}
where we factored out the dependence on the cutoff in each term of the sum. This formula is particularly suitable for discussing holographic renormalization. As we discussed above, our observer is a bulk object, as such, its physics depends on the distances in the bulk of the spacetime, and therefore on the holographic cutoff. We clearly see from the formula \eqref{eq:observerpropagatorCFT} that taking a strict $\epsilon \to 0$ limit in this setup results in the ground state of the quantum system being the only one that contributes to the propagator. If this is the case, then we get a finite, cutoff independent result by renormalizing with a power $\epsilon^{-2 \Delta_0}$. However, the fact that only the ground state contributes means that in this regime the observer is not affected by the gravitational coupling, since changing the Euclidean time evolution for a system that does not have any contribution from excited states does nothing. Contributions from other states are suppressed by $\epsilon^{2\alpha}$, with $\alpha= \sqrt{1/4+m^2+2m (E_0+\omega)}-\sqrt{1/4+m^2+2m E_0}$, where $\omega$ is the gap above the ground state. For large mass this reduces to $\epsilon^{2 \omega}$. Unlike in standard AdS/CFT, in this theory the cutoff is a physical quantity, related to a UV cutoff of the dual theory. We can therefore work in a regime where we still scale the cutoff to be small enough to ignore finite cutoff corrections to the action \ref{eq:JTaction}, but do not set it strictly to zero. If we do so, we can consider the gap $\omega$ to be small enough, in particular such that $\epsilon^{2 \omega}$ is not too small\footnote{This corresponds to $\log(1/\epsilon) \omega\sim 1$ or smaller, a bound which we will also encounter later on.}; then the observer picks up some of the gravitational effects. For the purposes of this work, we will work in this regime leave the above propagator unnormalized. In principle, it could be possible to keep the cutoff finite by improving the Schwarzian action to a finite cutoff one \cite{Iliesiu:2020zld, Stanford:2020qhm, Griguolo:2021wgy, Chaudhuri:2024yau, Griguolo:2025kpi}. It would be interesting to see how our result changes when pushing the wiggly boundary away from the actual AdS$_2$ boundary.

Since we are in a setup where the AdS/CFT correspondence is well established, we are also able to map the bulk quantity that we computed to a boundary calculation. The formula \eqref{eq:observerpropagatorCFT} allows us to do so. Each of the terms $\expval{\mathcal{O}^{\Delta_n}(0)\mathcal{O}^{\Delta_n}(u_1)}_{QG}$ can be mapped to observables in the dual SYK theory \cite{Maldacena:2016hyu}, even away from the conformal limit, or in the matrix model \cite{Jafferis:2022wez}. This gives a way to encode the observer's Euclidean propagator in the boundary theory. We can also go one step further and observe that the formula suggests that, in the dual theory, the observer manifests as a tower of operator insertions at some points of the thermal circle. It could be interesting to explore how to embed this in a theory of CFT with defects \cite{Billo:2016cpy}, which in this case would carry information about the observer. For the propagator, these defects must encode both the energy basis wavefunctions and the tower of dimensions $\Delta_n$.
\subsection{Semiclassical analysis}
In this section we discuss how the propagator computed above behaves in the semiclassical limit, working in a regime where the quantum system does not backreact on the worldline or on the boundary curve. This is also a regime where the results do not depend on the particular way we coupled the observer to the worldline, as we discuss in appendix \ref{app:differentactions}. As one expects from the classical limit, the measure of \eqref{eq:propagatormuintegral} localizes the integral over $\betaobs$ on a fixed value $\betaobs^*$, which is the geodesic distance between the two points on the boundary curve whose shape is determined by the classical equations of motion of JT gravity coupled to the worldline. We also characterize fluctuations around this value; namely, we derive a formula of the form
\begin{align}\label{eq:propagatormufluctuations}
	\bra{\psi_f} U_{QG}(\beta,u_1,m)\ket{\psi_i}=\int d\betaobs \exp{-\frac{\left(\betaobs-\betaobs^*\right)^2}{2\delta\betaobs^2}}\mel{\psi_f}{e^{-\betaobs H}}{\psi_i}\,,
\end{align}
where $\betaobs^*$ and $\delta \betaobs^2$ depend on $\beta,u_1,m$.
We will derive analytic formulae for both $\betaobs^*$ and the fluctuations $(\delta \betaobs)^2$ in semiclassical limits of weak and strong worldline backreaction, working in a regime where we neglect backreaction of the observer. One might worry that the explicit $\epsilon$ dependence of these expressions is only a matter of holographic normalization. The overall cutoff dependence of the boundary anchored propagator can indeed be removed by the renormalization discussed above. However, the cutoff also enters the saddle value and the width of the distribution of the proper time $\betaobs$, with $\betaobs^*\sim \log(1/\epsilon)$. From the point of view of the quantum system this dependence is physical, because $\betaobs$ is the Euclidean time appearing in the evolution operator $e^{-\betaobs H_{QM}}$. We therefore keep the unnormalized propagator and track its cutoff dependence explicitly. The variance then separates into a worldline QFT contribution (or Brownian term), $\delta\beta_{{\rm obs},{\rm wl}}^2\sim \betaobs^*/m\sim \log(1/\epsilon)/m$, and a genuine quantum gravity contribution that is cutoff independent at leading order, scaling as $\beta$ in the weak backreaction regime and as $1/m$ in the strong backreaction regime. Thus, for parametrically small $\epsilon$, the Brownian term dominates, and the relative size of the quantum gravity term is largest in the strong backreaction limit.

\subsubsection{Overview of the parameters and scaling}
Before turning to the calculation, let us explain in detail the parameters of the problem and the limits we are working in. Except for the IR cutoff $\epsilon \ll 1$, which sets the smallest scale in the problem, the quantity we computed depends on the following dimensionless parameters (in AdS units, remembering that the parameters $u_1,\beta$ are the rescaled ones and would include inverse powers of the renormalized dilaton $\beta/\phi_r, u_1/\phi_r$):
\begin{align}
	m \,, \quad u_1\,, \beta \,, \quad E_0\, \,,\quad \omega .
\end{align}
The last two parameters are the ground state energy of the quantum system, $E_0$, and the level spacing, $\omega$. As discussed in section \ref{sec:generalframework}, the semiclassical limit for the gravitational sector is achieved when $\beta,u_1$ are small (since $\phi_r$ is large). Another way to reach a semiclassical limit is to scale the mass of the worldline to be large, $m \gg 1$. Regarding the parameters of the quantum system, we will consider a situation where it is very weakly coupled to both gravity and the worldline so that we can ignore its backreaction. Expanding the states of the observer in the energy basis, we can write
\begin{align}\label{eq:QGpropagatorfinalenergybasis}
	&\bra{\psi_f} U_{QG}(\beta,u_1,m)\ket{\psi_i}=\int_0^\infty \frac{d\betaobs}{\betaobs^{3/2}} e^{-\frac{\betaobs m}{2}-\frac{\betaobs}{8m}} \int_0^\infty d\ell \,\ell\,\varphi_{u_1}(\ell)\varphi_{\beta-u_1}(\ell)\times\\
	&\times\int_{d(\ell)}^{\infty} \frac{dy \, y \, e^{-\frac{my^2}{2\betaobs}}}{\sqrt{\cosh y - \cosh d(\ell)}}\sum_{n} \psi_f^*(E_n)\psi_i(E_n)e^{-\betaobs E_0}e^{-\betaobs E_n(\omega)} \,,
\end{align}
Where $E_n(\omega)$ is the energy above the ground state and $\psi_i(E_n)$ are the wavefunctions in the energy basis. It is clear from this formula that we can ignore the backreaction of the observer as long as $E_n=E_0+E_n(\omega)$ is small compared to $m$ and to $1/\beta, 1/u_1$ (which enter the exponents). We can achieve this by for example setting a small ground state, a small level spacing $\omega$, and asking that the initial and finite states have small overlap with very large energy states.
Even though we do not consider the backreaction of the quantum system/observer, it is interesting to notice that the ground state energy of the quantum system is not irrelevant, as we expect when coupling a quantum system to gravity, since it enters the integrals in a different way than $m$. To summarize, ignoring the observer's backreaction, we can reach the semiclassical limit when either $1/\beta,1/u_1 \gg 1$ or $m \gg 1$, keeping their ratio fixed. The further scaling $m \beta \ll 1$ implies that backreaction of the worldline is negligible; this is the standard probe regime of JT gravity. For $m \beta \gg 1$, instead, we are in a semiclassical regime where the dynamics is dominated by worldline backreaction.

\subsubsection{Saddle point}\label{sec:saddlepoint}
We will first discuss the double scaling limit $m \gg 1$, $\beta \ll 1$ with $m \beta$ fixed, and then take further scaling limits to reach the two regimes described above. Since we are taking some parameters to be large, we can do the various integrals of \eqref{eq:QGpropagatorfinal} by saddle point. This is a five-dimensional saddle point problem; in principle, we should rescale the various coordinates to make the large parameter ($m$ or $1/\beta$) appear in front of the exponent, and write down the full system of saddle point equations. We will avoid rescaling the coordinates in order not to make the presentation too long; obviously this does not affect the results. Moreover, here the order of limits turns out not to be important: we get the same equations either by taking $m$ large or by taking $1/\beta$ large, always keeping $m\beta$ fixed. It is instructive to present the saddle point in the various variables one by one, and to clarify what it means from the geometric point of view. Let us start with the full expression, using the Wheeler-de Witt wavefunctions in the $\theta$ representation \eqref{eq:wdwtheta}
\begin{align}\label{eq:QGpropagatorfull}
	&K_{QG}=\frac{1}{(\beta-u_1)^{3/2}(u_1)^{3/2}}\int_0^\infty \frac{d\betaobs}{\betaobs^{3/2}} e^{-\frac{\betaobs m}{2}} \int_0^\infty \frac{d\ell}{\ell}\,\int_{d(\ell)}^{\infty} \frac{dy \, y \, e^{-\frac{my^2}{2\betaobs}}}{\sqrt{\cosh y - \cosh d(\ell)}}\times  \nonumber\\
	&\times \int_{c+i\mathbb{R}} d\theta_1 \, \theta_1\, e^{ \frac{\theta_1^2}{2u_1}+\frac{4}{\ell} \cos \theta_1/2} \int_{c+i\mathbb{R}} d\theta_2 \, \theta_2\, e^{ \frac{\theta_2^2}{2(\beta-u_1)}+\frac{4}{\ell} \cos \theta_2/2\,}\mel{\psi_f}{e^{-\betaobs H}}{\psi_i},
\end{align}
where we used the shorthand $\bra{\psi_f} U_{QG}(\beta,u_1,m)\ket{\psi_i}\equiv K_{QG}$. Starting from the integrals over $\betaobs$ (worldline proper time) and $y$ (fluctuations of the worldline away from the geodesic), the saddle point is at $\betaobs^*=y^*=d(\ell)$ (we ignore backreaction of the quantum system and drop prefactors and exponentials that do not scale with $1/\beta$ or $m$\footnote{We also keep the exponents proportional to $1/\ell$; we will see a posteriori that this indeed scales with one of the large parameters.}). This corresponds to the situation where the worldline is a geodesic in EAdS$_2$ of length $\betaobs^*=y^*=d(\ell)$, and the proper Euclidean time for which we evolve the observer degrees of freedom is also equal to this geodesic length. We are then left with the triple integral
\begin{align}\label{eq:QGpropagatorltheta}
	&K_{QG}=\frac{1}{(\beta-u_1)^{3/2}(u_1)^{3/2}}\int_0^\infty\frac{d\ell}{\ell} \frac{1}{ d(\ell)^{3/2}}\,e^{-d(\ell) m}  \nonumber\\
	&\times \int_{c+i\mathbb{R}} d\theta_1 \, \theta_1\, e^{ \frac{\theta_1^2}{2u_1}+\frac{4}{\ell} \cos \theta_1/2} \int_{c+i\mathbb{R}} d\theta_2 \, \theta_2\, e^{ \frac{\theta_2^2}{2(\beta-u_1)}+\frac{4}{\ell} \cos \theta_2/2\,}\mel{\psi_f}{e^{-d(\ell) H}}{\psi_i}\,.
\end{align}
We can think of the integral over $\ell,\theta_1,\theta_2$ as integrals over classical configurations of the boundary curve; see figure~\ref{fig:observer-lens-geometry}.
\begin{figure}
    \centering
    \includegraphics[width=0.58\linewidth]{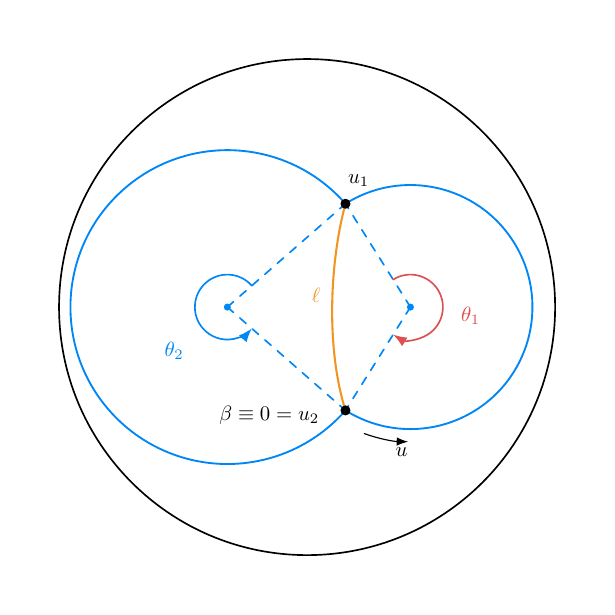}
    \caption{Classical boundary configurations contributing to the integral over $\ell,\theta_1,\theta_2$. The two blue arcs have common endpoints, the orange curve has length $\ell$, and the dashed radii indicate the angles $\theta_1,\theta_2$ associated with the two boundary segments.}
    \label{fig:observer-lens-geometry}
\end{figure}
As we have discussed before, the WdW wavefunctions we used are accurate only if $\ell/\epsilon \gg 1$, so we can use this limit to simplify $d(\ell)\simeq 2 \log(\frac{\ell}{\epsilon})$. Then the $\ell$ saddle point gives
\begin{align}
	\ell^*(\theta_1,\theta_2)=-\frac{2}{m}(\cos(\theta_1/2)+\cos(\theta_2/2))\,,
\end{align}
where we do not take into account powers of $\ell$ in the prefactor that do not depend on the large parameters. The saddle point equations for the $\theta_{1,2}$ integrals are
\begin{align}
	&\theta_{1} +
mu_1\dfrac{ \, \sin\!\left(\tfrac{\theta_{1}}{2}\right)}
     {\cos\!\left(\tfrac{\theta_{1}}{2}\right) + \cos\!\left(\tfrac{\theta_{2}}{2}\right)} = 0, \\
     &\theta_{2} +
mu_2\dfrac{ \, \sin\!\left(\tfrac{\theta_{2}}{2}\right)}
     {\cos\!\left(\tfrac{\theta_{1}}{2}\right) + \cos\!\left(\tfrac{\theta_{2}}{2}\right)} = 0
\end{align}
where $u_2\equiv\beta-u_1$. Notice that these two equations only depend on the dimensionless combinations of parameters that we are keeping fixed, $m u_{i}$, $i=1,2$. The system always has a solution $\theta_1=\theta_2=0$, which gives negative $\ell$, and we discard it\footnote{Including prefactors of the type $\ell^{\alpha}$ with $\alpha$ independent of $m, \phi_b$ would move this saddle point away from zero, but it would still give negative $\ell$.}. It is possible to check numerically that, for any value of the parameters, there is another solution with $\theta_{1,2}\in (\pi,2 \pi)$, yielding positive $\ell$. These equations can be solved analytically in the further limits $m u_i\gg 1$ and $m u_i\ll 1$, corresponding to strong and weak backreaction. In the first limit, we get
\begin{align}\label{eq:solutionslargembeta}
&\theta_{1}^* =2 \pi  -\frac{8\pi}{mu_1}+ \order{\left(\frac{1}{mu_1}\right)^{2}}, \\
&\theta_{2}^* = 2 \pi -\frac{8\pi}{mu_2} + \order{\left(\frac{1}{mu_2}\right)^{2}},\\
&\ell^*=\frac{4}{m}\left[
1-4 \pi^2 \left(\left(\frac{1}{mu_2} \right)^2
    + \left(\frac{1}{mu_1}\right)^2 +\order{(m u_i)^{-3}}
\right)\right]\,.
\end{align}
The other saddle points are then
\begin{align}
	\betaobs^*=y^*=2 \log(\frac{4}{m \epsilon})-\frac{8\pi^2\left(u_1^2+u_2^2\right)}{u_1^2u_2^2 m^2}
\end{align}
These saddles are close to $\theta_{1,2}=2\pi$. We can understand this as the regime where the geometry that dominates is the one in the left panel of figure~\ref{fig:observer-lens-angle-regimes}, where backreaction of the worldline is very large. Notice that in order to keep $\ell^*/\epsilon \gg 1$, there is an upper bound on the mass: $m \epsilon \ll 1$, as we had anticipated in section \ref{sec:generalframework}.

In the other limit, $m u_i \ll 1$, we instead get the following saddle points for the angles:
\begin{align}\label{eq:solutionssmallmbeta_theta}
\theta_1^* &= \frac{2\pi u_1}{\beta}
+ mu_1 \left( \frac{1}{\pi}
+ \frac{\beta-u_1}{\beta} \cot\!\left(\tfrac{\pi u_1}{\beta}\right) \right) \\
&\quad + \frac{m^2u_1}{4\pi^3}\left(-2\beta-\pi(\beta-2u_1)
   \cot\!\left(\tfrac{\pi u_1}{\beta}\right)\right) \\[6pt]
\theta_2^* &= \frac{2\pi(\beta-u_1)}{\beta}
+ 	m(\beta-u_1) \left( \frac{1}{\pi}
- \frac{u_1}{\beta} \cot\!\left(\tfrac{\pi u_1}{\beta}\right) \right) \\
&\quad - \frac{m^2(\beta-u_1)}{4\pi^3}\left(2\beta+\pi(\beta-2u_1)
   \cot\!\left(\tfrac{\pi u_1}{\beta}\right)\right).
\end{align}
Notice that we need these to order $\left(m\beta\right)^2$ if we want $\ell$ at first order. This gives the following saddle point value of $\ell$:
\begin{align}\label{eq:solutionssmallmbeta_l}
\ell^*&=\frac{\beta\,\sin\!\left(\tfrac{\pi u_1}{\beta}\right)}{\pi}-\\
&- \frac{m}{2\pi^3} \Bigg[
   \pi \cos\!\left(\tfrac{\pi u_1}{\beta}\right)
   \left( \beta(\beta-2u_1)
   + \pi u_1(-\beta+u_1)
   \cot\!\left(\tfrac{\pi u_1}{\beta}\right) \right)
   \;+\; \beta^2 \sin\!\left(\tfrac{\pi u_1}{\beta}\right)
\Bigg],
\end{align}
This reduces to $\ell^*=\frac{\beta}{\pi}-\frac{m\beta^2}{2\pi^3}$ for the symmetric case $u_1=u_2=\beta/2$. Notice that in this case the requirement $\ell^*/\epsilon \gg 1$ implies\footnote{If one restores $\phi_r$, the condition is $\epsilon \phi_r /\beta \ll 1$, which is stronger than the condition that corrections to the Schwarzian action are subleading: $\beta/\epsilon \ll1$.} $\epsilon/\beta \ll1$, i.e. the effective Planck length cannot be smaller than the cutoff. The on shell value of $\betaobs$ is instead
\begin{align}
	&\betaobs^*=y^*=2 \log(\frac{1}{\epsilon}\frac{\beta\,\sin\!\left(\tfrac{\pi u_1}{\beta}\right)}{\pi})- \\
	&- \frac{2\pi}{\beta\,\sin\!\left(\tfrac{\pi u_1}{\beta}\right)}\frac{m}{2\pi^3} \Bigg[
   \pi \cos\!\left(\tfrac{\pi u_1}{\beta}\right)
   \left( \beta(\beta-2u_1)
   + \pi u_1(-\beta+u_1)
   \cot\!\left(\tfrac{\pi u_1}{\beta}\right) \right)
   + \beta^2 \sin\!\left(\tfrac{\pi u_1}{\beta}\right)
\Bigg]+\nonumber\\
&+\order{\left(m\beta\right)^2}\,.\nonumber
\end{align}
For the symmetric case this is just
\begin{align}
	\betaobs^*=d(\ell^*)= 2 \log(\frac{\beta}{\pi\epsilon})-\frac{m\beta}{\pi^2}+\order{\left(m\beta\right)^2}
\end{align}
These saddle points correspond to the situation depicted in the right panel of figure~\ref{fig:observer-lens-angle-regimes}, where the boundary curve is almost a perfect disk and the worldline backreacts only mildly.
\begin{figure}
    \centering
    \subfigure[$m u_i\gg1$: $\theta_{1,2}\simeq2\pi$.]{%
        \includegraphics[width=0.48\linewidth]{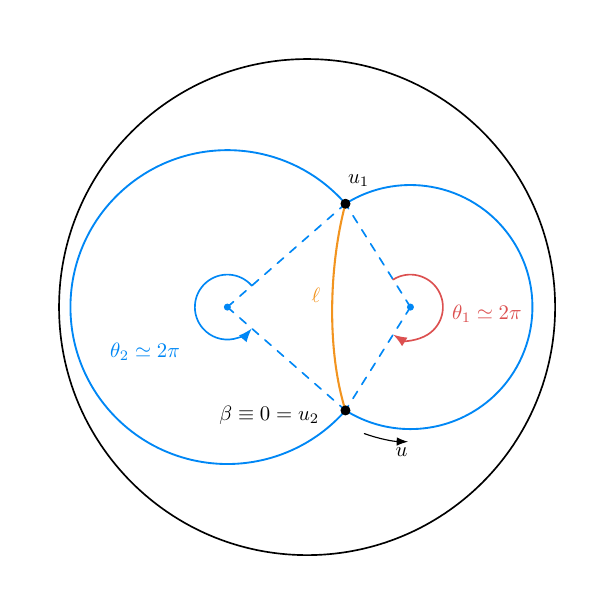}
    }
    \hfill
    \subfigure[$m u_i\ll1$: disk-like boundary curve.]{%
        \includegraphics[width=0.48\linewidth]{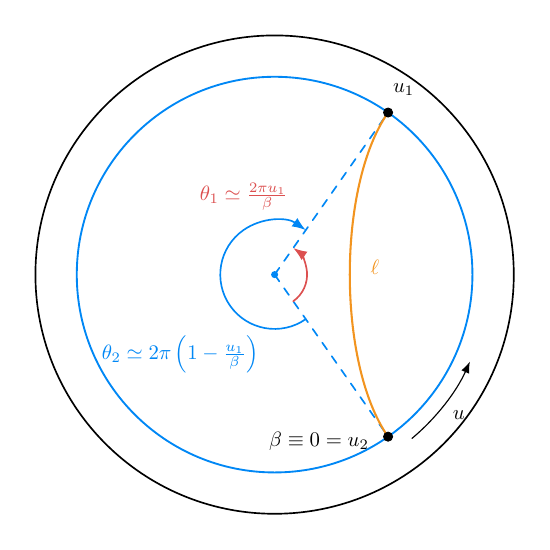}
    }
    \caption{Classical geometries at the two saddle point regimes. In the large-backreaction regime the two Schwarzian boundaries are joined with a strong kink and the two angles sit near $2\pi$. In the mild-backreaction regime the boundary curve is close to a circle and the angles are set by the fractional boundary time separations.}
    \label{fig:observer-lens-angle-regimes}
\end{figure}

The saddle points we just found correspond to the classical solutions for the gravitational boundary mode coupled to the worldline. The total Euclidean time $\betaobs$, collapses to the geodesic length stretching between the two boundary points separated by a renormalized distance $u_1$ on the boundary curve. We obtain the same results by solving the equations of motion of the theory in appendix \ref{app:EOM}. Thus we see that, in the semiclassical limit, the quantum theory of the observer is again described by a single evolution operator at a fixed Euclidean time $U_{QM}(\betaobs^*)$, where $\betaobs^*$ is determined by the equations of motion.

\subsubsection{Fluctuations}\label{sec:fluctuations}
Having analyzed the saddle point, we now characterize fluctuations around it, deriving the spread $\delta\betaobs^2(\beta,u_1,m)$. Since we are mostly interested in the scaling of the fluctuations with the parameters of the theory that govern the semiclassical regime, we will work for simplicity in the symmetric case $u_1=\beta/2$. The integral for the quantum mechanical propagator that we should study is
\begin{align}\label{eq:QGpropagatorfinal2}
	&\bra{\psi_f} U_{QG}(\beta,u_1=\beta/2,m)\ket{\psi_i}=\int_0^\infty \frac{d\betaobs}{\betaobs^{3/2}} e^{-\frac{\betaobs m}{2}-\frac{\betaobs}{8 m}} \int_0^\infty d\ell \,\ell\,\varphi_{\beta/2}(\ell)\varphi_{\beta/2}(\ell)\times\\
	&\times\int_{d(\ell)}^{\infty} \frac{dy \, y \, e^{-\frac{my^2}{2\betaobs}}}{\sqrt{\cosh y - \cosh d(\ell)}} \mel{\psi_f}{e^{-\betaobs H}}{\psi_i}\,.
\end{align}
As in the saddle point analysis, we will neglect the backreaction of the quantum system. In the semiclassical limit, we scale both $m$ and $\frac{1}{\beta}$ to be large. After writing the WdW wavefunctions using the $\theta$ representation, we are left with the five-dimensional integral described in the previous section. The physical quantity from the point of view of the observer is the proper time $\betaobs$, so we are interested in calculating the spread of the distribution $\mu(\betaobs)$ in the $\betaobs$ direction, which we call $\Delta \betaobs=\betaobs-\betaobs^*$. It turns out that fluctuations of $y$ do not matter in the calculation of this spread\footnote{To see this, consider the two-dimensional integral
\begin{align}
	\int_d^\infty d y \int_0^\infty d\betaobs e^{-\frac{\betaobs m}{2}} \frac{e^{-\frac{m y^2}{2\betaobs}}}{\sqrt{\cosh y-\cosh d}}\,.
\end{align}
If we expand it around the saddle $y=\betaobs=d(\ell(\theta))$ we have
\begin{align}
	e^{-m d}\int_0^\infty d( \sqrt{\Delta y})e^{-m (\sqrt \Delta y)^2} \int_{-\infty}^\infty d\betaobs e^{-\frac{(\Delta \betaobs)^2 m}{2 d(\ell)}}
\end{align}
The $\sqrt{\Delta y}$ integral can be extended over $\mathbb R$, so these are now two Gaussians and there is no cross term between $\betaobs$ and $y$. Moreover, the spread of the $\sqrt{\Delta y}$ Gaussian does not depend on $\ell,\theta_i$. This means that we can compute expectation values of $(\Delta \betaobs)^2$ (or also of $\Delta \ell^2, \Delta\theta^2$) without worrying about fluctuations of $y$, simply because they do not mix. This remains true even if we consider contributions of the form $e^{-\betaobs E_n}$ from the observer.}. This also implies that at large mass, for these fluctuations, the ground state energy of the quantum system can be reabsorbed by a shift in the mass. We can simply consider the integral over $y$ by saddle point, which fixes $y=d(\ell)$, and ignore its fluctuations. This is the same answer that one would get by taking them into consideration when expanding the full five-dimensional saddle point. Apart from this, keeping track of the exponents that scale with the large parameters, we have
\begin{align}\label{eq:muintegralpresaddle}
	&\int d\betaobs \mu(\betaobs)\mel{\psi_f}{e^{-\betaobs H_{QM}}}{\psi_i}\nonumber\\
	&\qquad=\int d\betaobs \, d\ell \, d\theta_1 \, d\theta_2\,
	\mathcal{N}_{\betaobs,\ell, \theta_1,\theta_2}
	e^{S(\betaobs,\ell, \theta_1,\theta_2)}
	\mel{\psi_f}{e^{-\betaobs H_{QM}}}{\psi_i}\,,\\
	&\text{with }S=-\frac{\betaobs m}{2}-\frac{m d(\ell)^2}{2\betaobs}+\frac{\theta_1^2+\theta_2^2}{\beta}+\frac{4}{\ell}\left(\cos \frac{\theta_1}{2}+\cos \frac{\theta_2}{2}\right)\,.
\end{align}
Expanding the integral at quadratic order in the fluctuations around the saddle point values
\begin{align}
	\int d\Delta x_i \exp{\Delta x_i \Delta x_j \partial_{ij} S(x_i^*)}
\end{align}
where $x_i=\{\betaobs,\ell, \theta_1,\theta_2\}$, and $\Delta x_i= x_i-x_i^*$ are the deviations from the saddle point values. Then, integrating out all the variables (along the steepest descent direction) except $\Delta \betaobs$, we are left with an expression of the form \eqref{eq:propagatormufluctuations}, with
\begin{align}\label{eq:deltabAbbetc}
	\frac{1}{(\delta \betaobs)^2}=|A_{\betaobs\betaobs}|-\frac{|A_{\ell\betaobs}|^2}{|A_{\ell\ell}|+2 \frac{|A_{\ell\theta_1}|^2}{|A_{\theta_1 \theta_1}|}}
\end{align}
where we have defined $A_{ij}=\partial_{ij} S(x_i^*)$. It can be shown that the first term is always of the form $|A_{\betaobs\betaobs}|=\frac{m}{d(\ell^*)}$. This term does not come from quantum gravity effects, but only from the fact that the worldline is allowed to fluctuate in the rigid bulk of the spacetime: it parameterizes quantum fluctuations of the worldline, we would have this term even if the boundary mode is frozen. The matrix $A_{ij}$ can be expressed analytically in the regimes of weak backreaction $m\beta \ll 1$ or strong backreaction $m\beta \gg 1$. The full expressions are cumbersome, so we will not write them here. The only important note is that the absolute values we reported above are selected by the steepest descent direction of each integral $\Delta x_i$. We will just quote the final result for the variance $(\delta \betaobs)^2$. In the first limit, we get
\begin{align}
	\delta\betaobs^2(\beta,m)=\frac{ 2 \log \left(\frac{\beta }{\pi  \epsilon }\right)}{m}+\frac{\beta }{\pi ^2}+\frac{\left(\pi ^2-20\right) \beta ^2 m}{8 \pi ^4}+ \order{\beta^3 m^2}.
\end{align}
In this case we have $\betaobs^*=d(\ell^*)= 2 \log(\frac{\beta}{\pi\epsilon})$, and we see that the first term is exactly of the form $\frac{d(\ell^*)}{m}$ and comes from the first term in \eqref{eq:deltabAbbetc}. This contribution should not be understood as a quantum gravity effect, since it comes from the fluctuations of the worldline only and gives the same scaling as Brownian type fluctuations. The second term is instead a genuine quantum gravity effect, and indeed scales with the two-dimensional analog of the Planck length (restoring $\phi_r$ and $r_{\rm{AdS}}$ we obtain the contribution $\delta \beta_{\rm{obs},QG}$ quoted in the introduction). Since the second is the term that we would like to capture, we could think of increasing the mass or $\beta$ to make it dominant, however since the term $\log(\pi \beta/\epsilon)$ should be large to stay in the infinite cutoff regime we must have $m \beta \gg 1$, entering the strong backreaction regime. In this limit we instead get
\begin{align} \label{eq:bfluct_largembeta}
	\delta\betaobs^2(\beta,m)=\frac{2 \log \left(\frac{4}{m \epsilon }\right)}{m}+\frac{2}{m}-\frac{128\pi^2}{\beta^2 m^3}+ \order{ \frac{1}{m^4 \beta^3}}
\end{align}
The first term is again the Brownian term. Notice that the quantum gravity contribution scales with the inverse mass, because the boundary mode fluctuations are fully governed by backreaction. In both cases we analyzed, the fluctuations include a term due only to the fluctuations of the worldline, i.e. this Brownian motion term:
\begin{align}
	\delta\betaobs^2(\beta,m)=\frac{\betaobs^*}{m }\,.
\end{align}
In addition to this contribution, one has a genuine, length independent quantum gravity contribution, which in the two limits scales with $\beta$ or $1/m$. As discussed above, the Brownian term is enhanced by a logarithmic factor of the IR cutoff, $\log(1/\epsilon)$. However, we would like to understand the regime where the genuine quantum gravity corrections are largest. From the analytic formulae we derived, it is clear that for sufficiently small values of $\epsilon$, such that $\betaobs^*$ has the same scaling in both limits we considered, the largest quantum gravity contribution (relative to this Brownian motion term) comes from the strong backreaction limit. We have also checked, by solving the equations numerically for generic $m\beta$, that the ratio of the Brownian term to the quantum gravity one is largest in such limit, where numerics agrees with the analytic formulae above. Let us now consider the relative fluctuations with respect to the saddle. The leading, Brownian term scales as
\begin{align}
	\frac{\delta\betaobs^2(\beta,m)}{\left(\betaobs^*(\beta,m)\right)^2}=\frac{1}{m\betaobs^*}\simeq \frac{1}{2 \log(\frac{1}{\epsilon})m}\ll 1\,,
\end{align}
and the quantum gravity contributions are further suppressed by powers of $\log(1/\epsilon)$. One can also consider the exact measure and characterize the relative variance of such distribution, as we describe in section \ref{sec:nonpertvariance}. We find that this relative variance is still largest in the semiclassical limit.

We therefore conclude that the relative fluctuations are largest in the semiclassical regime of large backreaction, where they are given by \eqref{eq:bfluct_largembeta}, but still suppressed by the inverse logarithm of the IR cutoff.

One might still expect to the observer to experience such fluctuations, as we will discuss momentarily.

\subsubsection{Can the observer feel quantum gravity?} \label{sec:pickup}
So far we have considered the relative fluctuations $\delta \betaobs/\betaobs^*$; however, one can compare the fluctuations with the level spacing of the quantum system, which we denote with $\omega$. If this ratio is close to one, then it means that the fluctuations of Euclidean time can induce transition between nearby energy levels, and the observer will feel the effect of such fluctuations. We will show that, for small level spacing, this happens even if $\delta \betaobs/\betaobs^*$ is small. 

Let us consider a particular observable, such as the average energy of the system. In ordinary thermodynamics, 
this is given by
\begin{align}
	\expval{E}_{th}^{(\betaobs)}=-\partial_{\betaobs} \log (Z_{QM}(\betaobs))\,.
\end{align}
Here, the Euclidean time $\betaobs$ should be interpreted as the genuine inverse temperature $\betaobs=1/T$. With quantum gravity, the inverse temperature fluctuates. In particular, we can use the formulae we derived earlier to define a quantum gravity partition by integrating the propagator between two fixed energy eigenstates against the density of states $\rho(E)$ of the observer:
\begin{align}
	Z_{QG}
	&= \int dE \rho(E) \mel{E}{U_{QG}}{E}
	= \int dE \rho(E) \int d\betaobs \, \mu(\betaobs)e^{-\betaobs E}\nonumber\\
	&= \int d\betaobs \, \mu(\betaobs) Z_{QM}(\betaobs)\,,
\end{align}
where $Z_{QM}(\betaobs)$ is nothing but the standard partition function of the quantum system at inverse temperature $\betaobs$. We can define the quantum gravity average of the energy as
\begin{align}
	\expval{E}_{QG}=\frac{1}{\int d\betaobs \mu(\betaobs)}\int d\betaobs \,\mu(\betaobs)\expval{E}_{th}^{(\betaobs)}\,.
\end{align}
Let us test this proposal with a simple system that only has the ground state. In this case $\expval{E}_{th}^{(\betaobs)}=E_0$ and is independent of $\betaobs$, and therefore we have
\begin{align}\label{eq:averagegroundstate}
	\expval{E}_{th}^{(\betaobs)}=\expval{E}_{QG}\,,
\end{align}
as one expects for such a simple system, since fluctuations of the temperature will not induce changes in states when there are no states to move to. In the semiclassical limit, $\mu(\betaobs)$ \eqref{eq:propagatormufluctuations} is a sharply peaked distribution around the classical value $\betaobs^*$, we can expand the energy around this value when $\delta \betaobs^*/\betaobs^* \ll 1$, getting
\begin{align}
	\expval{E}_{QG}=\expval{E}_{th}^{(\betaobs^*)}+ \frac{1}{2}(\delta \betaobs)^2 \partial_{\betaobs}^2 E(\betaobs^*)+\dots .
\end{align}
Close to the saddle point we have $\partial_{\betaobs}^2 E(\betaobs^*)= -\partial_{\betaobs}^3 \log(Z_{QM}(\betaobs))$. Now let us consider a particular case of a bosonic partition function with a ground state $E_0$ and a gap $\omega$
\begin{align}
	Z_{QM}(\betaobs)=\frac{e^{-E_0 \betaobs}}{1-e^{-\betaobs \omega }}
\end{align}
Then for small $\omega \betaobs^*$, i.e. small level spacing with respect to the on shell value of the total Euclidean time, one gets (up to numerical factors, and after substituting the semiclassical $(\delta \betaobs)^2$)
\begin{align}
	\delta E_{QG}\equiv \expval{E}_{QG}-\expval{E}_{th}^{(\betaobs^*)} = \frac{1}{m (\betaobs^*)^2}+ \frac{\delta\beta_{{\rm obs},QG}^2}{(\betaobs^*)^3},
\end{align}
where we denoted $\delta\beta_{{\rm obs},QG}^2$ the quantum gravity contribution to the variance, which is $\delta\beta_{{\rm obs},QG}^2=2/m$ in the strong backreaction regime and $\delta\beta_{{\rm obs},QG}^2=\beta/\pi^2$ in the weak backreaction regime. We can compare this difference with the level spacing of the system, if their ratio is of order one, then it means that the fluctuations can affect the quantum system by inducing transitions between nearby states. The ratio is, in the strong and weak worldline backreaction limits
\begin{align}
	&\frac{\delta E_{QG}}{\omega}= \frac{1}{\omega \betaobs^*}\frac{1}{m \betaobs^*} \left(1+\frac{1}{\betaobs^*}\right)\,,\quad m \beta \gg 1\\
	&\frac{\delta E_{QG}}{\omega}= \frac{1}{\omega \betaobs^*}\frac{1}{m \betaobs^*} + \frac{1}{\omega \betaobs^*} \frac{2 \pi \beta}{(\betaobs^*)^2}\,,\quad m \beta \ll 1.
\end{align}
This shows that the observer is sensitive to Brownian-type fluctuations if $\omega \sim \frac{1}{m (\betaobs^*)^2}$, and even to to genuine gravitational effects if $\omega \sim \frac{1}{m (\betaobs^*)^3}$ or $\omega \sim \frac{\beta}{ (\betaobs^*)^3}$. This requires the condition $\omega \log(1/\epsilon) \ll 1$  (or stronger), which is compatible with the scaling needed to obtain contributions to the propagator from states beyond the ground state, as discussed in \ref{sec:renormalization}. Since there is no fundamental obstruction to taking this level spacing arbitrarily small, we can in principle make these fluctuations comparable to the level spacing. Thus, even when the relative fluctuations $\delta \betaobs/\betaobs^*$ are small, the observer will feel them.

This is also a good point of connection with works that use the observer as a clock to dress gravitational observables to the worldline \cite{Witten:2023background,Chandrasekaran:2022cip, Witten:2021unn, Mirbabayi:2023vgl}, and in particular  \cite{Kolchmeyer:2024fly, Yang:2025lme}, where the observer/clock is treated at the quantum level neglecting gravitational backreaction. If one uses our observer as clock, the regime we just discussed, where the clock is sensitive to quantum gravity fluctuations, will also be the same where a proper quantum treatment of the clock is necessary. It would be interesting to develop the connection with the works \cite{Kolchmeyer:2024fly, Yang:2025lme} further, since in our treatment we also allow interactions of the observer/clock with gravity.

\subsection{Non perturbative variance} \label{sec:nonpertvariance}
Till now we discussed physics close to the semiclassical limit, where we can trust the saddle point approximation. However, since we have an exact formula for the propagator, one can also look at the full distribution $\mu(\betaobs)$ and compute the variance of $\betaobs$ away from the semiclassical limit. This allows us to see whether and how the fluctuations grow larger away from the semiclassical limit, claryifying up to which point we can relax the assumption on the small level spacing we needed to for the observer capture quantum gravitational effects. We can characterize fluctuations induced by the measure (still ignoring backreaction of the quantum system) using the variance
\begin{align}
	\mathrm{Var}_{full}= \expval{(\betaobs- \expval{\betaobs})^2}\,, \quad \expval{f(\betaobs)}=\frac{\int d\betaobs \mu(\betaobs) f(\betaobs)}{\int d\betaobs \mu(\betaobs)}\,.
\end{align}
The first thing we can look at is the relative variance with respect to the average value $\expval{\betaobs}^2$, i.e. we look at the quantity
\begin{align}
	 \frac{\expval{(\betaobs- \expval{\betaobs})^2}}{\expval{\betaobs}^2}\,.
\end{align}
By plotting this function as a function of $1/\beta$ or $m$, one sees that the average value $\expval{\betaobs}^2$ grows faster than the variance $\expval{(\betaobs- \expval{\betaobs})^2}$ when moving away from the semiclassical limit. As a result, the largest relative fluctuation still comes from the semiclassical regime, as we anticipated in section \ref{sec:fluctuations}. Since this does not add new information, we will not report the plots here.

\begin{figure}
    \centering

    \begin{minipage}{0.48\textwidth}
        \centering
        \IfFileExists{figures/full_frozen_corrected_eps001_beta10_m_slice_m20_full_variance.pdf}
        {\includegraphics[width=\linewidth]{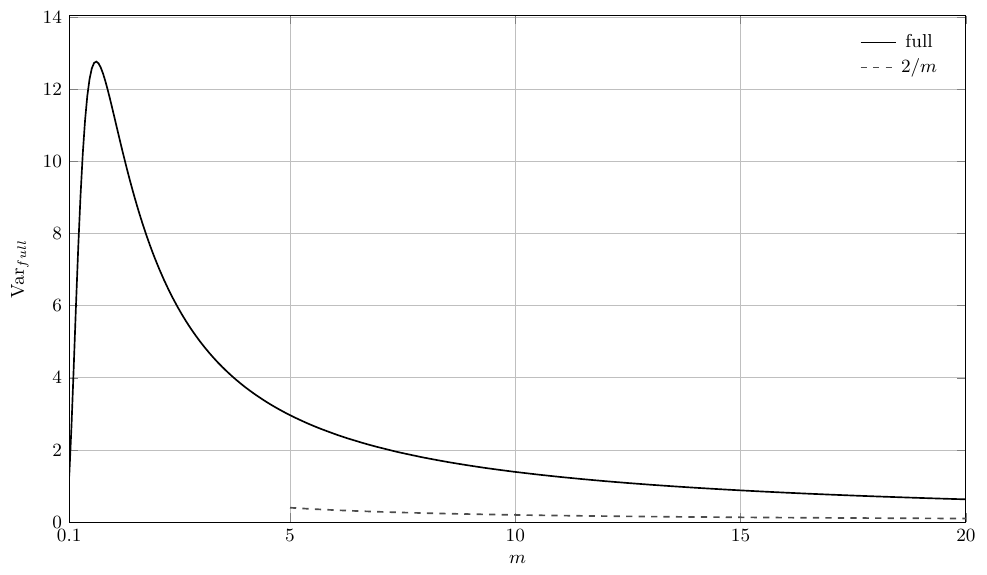}}
        {\rule{\linewidth}{3cm}}
        \small (a)
        \label{fig:plot13}
    \end{minipage}
    \hfill
    \begin{minipage}{0.48\textwidth}
        \centering
        \IfFileExists{figures/full_frozen_corrected_eps001_beta10_m_slice_m20_variance_difference.pdf}
        {\includegraphics[width=\linewidth]{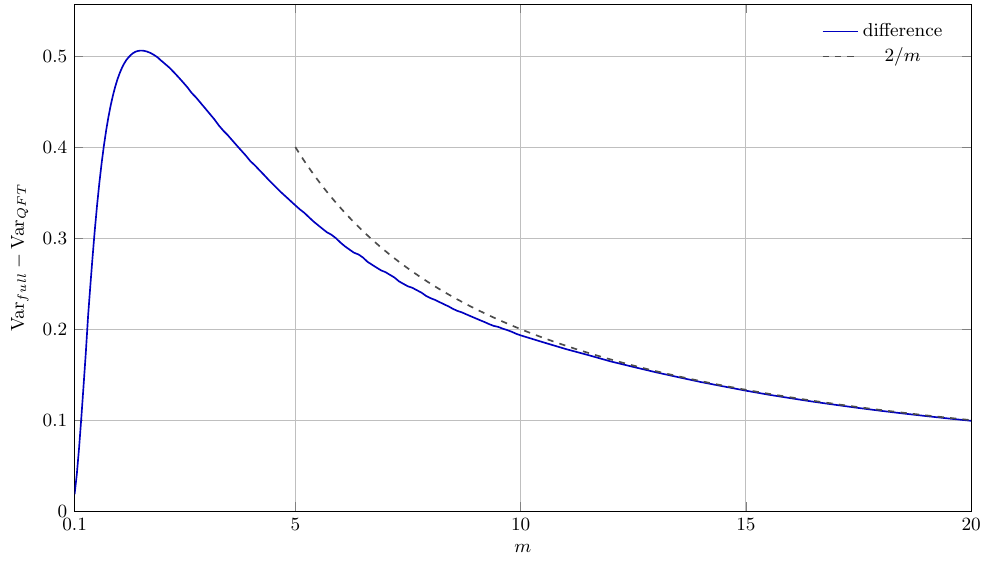}}
        {\rule{\linewidth}{3cm}}
        \small (b)
        \label{fig:plot23}
    \end{minipage}

    \caption{Variance plots for $\beta=10$, $\epsilon=10^{-3}$. a) Full variance plot. b) Difference between the full variance and the QFT variance, which should only come from quantum gravity. The dotted line is the semiclassical approximation of the quantum gravity variance in the large $m\beta$ regime, and it agrees with numerics.}
    \label{fig:nonpertfixedbeta}
\end{figure}

Since we have discussed an example where the variance must be compared with a scale different from the average value, we can also examine the variance in isolation. To disentangle the contributions to the variance from QFT/Brownian motion and quantum gravity, as we did in the semiclassical case, we plot both the full variance and another quantity defined as
\begin{align}
	\mathrm{Var}_{QG}= \mathrm{Var}_{full}-\mathrm{Var}_{QFT}\,,
\end{align}
where $\mathrm{Var}_{QFT}$ is the variance computed by fixing the gravitational variables $\theta_1,\theta_2,\ell$ to their saddle point values and computing the variance in $\betaobs$ from only the QFT part of the effective action (the $y,\betaobs$ integral). The results for $\beta=10$ (away from the semiclassical regime) and $\epsilon=10^{-3}$ are shown in figure \ref{fig:nonpertfixedbeta} as a function of the mass of the worldline. We see that the difference agrees with the semiclassical quantum gravity contribution at large mass. Moving away from the large mass semiclassical regime, it grows less than the semiclassical extrapolation; however, the maximum occurs at a value that is not parametrically small. The turning point and decrease of the absolute variance at very small mass should not be interpreted as a recovery of semiclassicality. Rather, this is the regime where the approximation that the observer is localized on a worldline breaks down; in fact this regime is also where the choice of the observer action we used is most relevant. Moreover, we can check the expectation that increasing $\beta$ moves us further from the semiclassical regime and increases fluctuations. This is shown in the three-dimensional plot in figure \ref{fig:3dplot}, where we plot the variance as a function of $m$ and $1/\beta$. 

Even though we still see that the QFT fluctuations are larger than the quantum gravity ones, these non perturbative analysis shows that quantum gravity effects can grow larger away from the semiclassical limit, and not be parametrically suppressed even when the cutoff is very small. This implies that they can be detected more easily by the observer, with less strong assumptions on its level spacing. These effects could be increasingly relevant when taking into account higher topologies, the analysis of such contribution will be discussed in a separate work.

\begin{figure}
    \centering

    \begin{minipage}{0.48\textwidth}
        \centering
        \includegraphics[width=\linewidth]{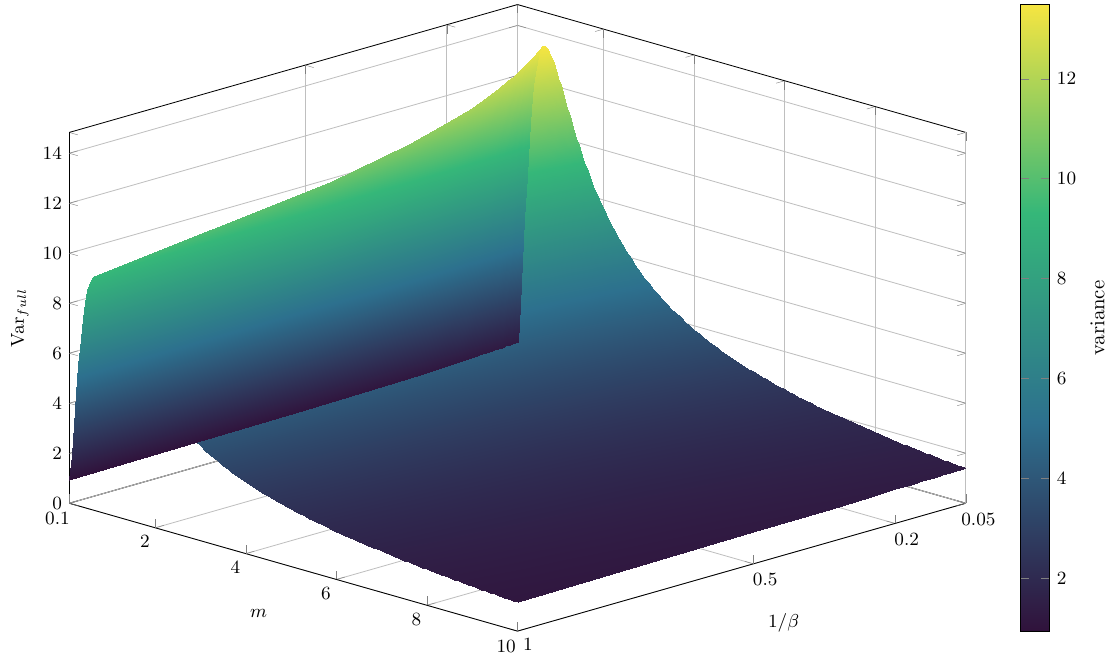}
        \small (a) Full variance plot
        \label{fig:plot1}
    \end{minipage}
    \hfill
    \begin{minipage}{0.48\textwidth}
        \centering
        \includegraphics[width=\linewidth]{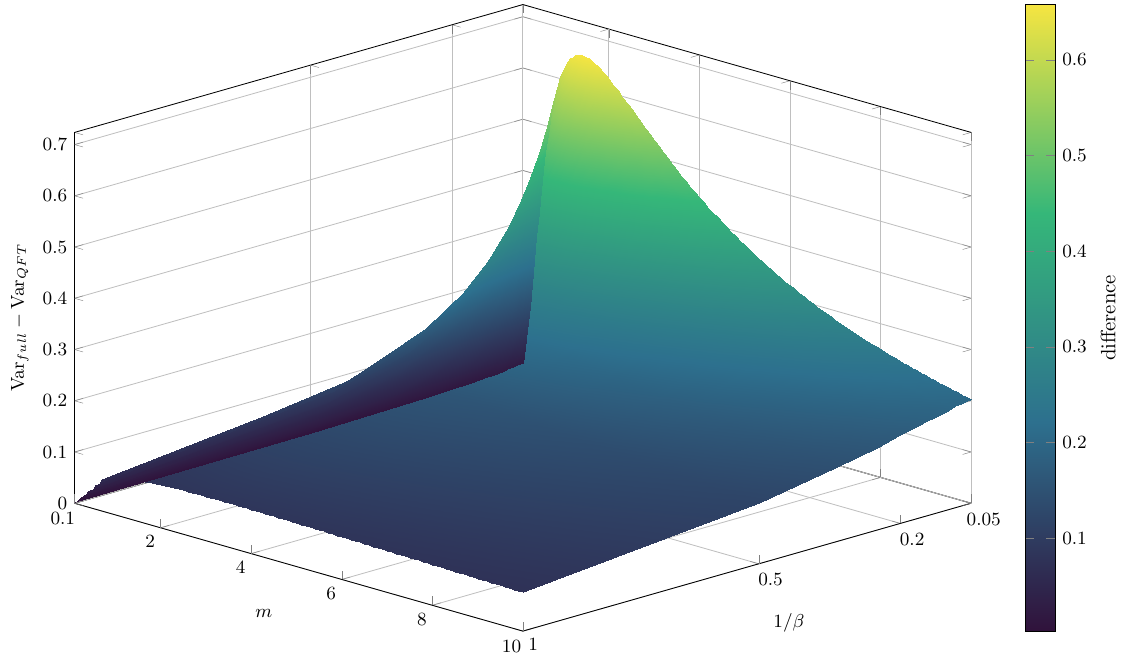}
        \small (b) Difference plot
        \label{fig:plot3}
    \end{minipage}

    \caption{Variance plots for $\epsilon=10^{-3}$ and varying $\beta$. As we expect, increasing $\beta$ moves us further from the semiclassical regime and increases fluctuations. a) Full variance plot. b) Difference between the full variance and the QFT variance, only comes from quantum gravity.}
    \label{fig:3dplot}
\end{figure}

\subsection{Lorentzian interpretation}
In the previous subsections we analyzed the Euclidean propagator. To relate it to the physics of a Lorentzian observer, we can analytically continue the results obtained above. The quantum gravity Euclidean propagator we derived is closely related to gravitationally dressed correlation functions on the Schwarzian boundary. For such observables, the continuation to the same boundary of the Lorentzian thermal AdS$_2$ geometry (i.e. the two dimensional eternal black hole) is obtained by taking\footnote{A different analytic continuation brings the start and end of the evolution of the quantum system to the two sides of the TFD state, i.e. to the two sides of the eternal black hole. This can be achieved by taking $u \to it+ \beta/2$. With such an analytic continuation, one can get a semiclassical limit that is a spacelike geodesic connecting the two points, i.e. an ER bridge, on top of which the observer lives. Although this analytic continuation is interesting for studying observables like the length of the ER bridge, it is not very useful for discussing the evolution of an observer coupled to quantum gravity, so we will not discuss it.} $u \to \epsilon+i t$, with $t=t_1-t_2$, where the sign of $\epsilon$ fixes the operator ordering. We can perform an analogous continuation for our Euclidean propagator \eqref{eq:QGpropagatorfinal}:
\begin{align}
	\mel{\psi_f}{U_{QG}(u_1=i(t_1-t_2)+\epsilon)}{\psi_i}=\mel{\psi_f}{U_{QG}(t_1-t_2)}{\psi_i}=\braket{\psi_f(t_2)}{\psi_i(t_1)}\,.
\end{align}
We would like to interpret this quantity as an overlap of states prepared at the Lorentzian boundary at times $t_1$ and $t_2$. It can be easily checked that it has the expected conjugation properties. One can also deform the $\betaobs$ contour in \eqref{eq:QGpropagatorfinal} to $\betaobs=i \tau(1-i \varepsilon)$, where $\varepsilon$ keeps the $\tau$ integral convergent, obtaining
\begin{align}\label{eq:overlaptauintegral}
	\braket{\psi_f(t_2)}{\psi_i(t_1)}=\int_0^\infty d\tau \mu(\tau)\mel{\psi_f}{e^{-i \tau H}}{\psi_i}
\end{align}
where $\mu(\tau)$ depends on the parameters of the theory and on $t_1-t_2$. Let us discuss a few points on unitarity. This formula casts doubt on the expectation that an observer should experience unitary evolution when coupled to quantum gravity, which is recovered in the semiclassical limit where the measure localizes the integral on a single value of $\tau^*$. However, we should also be able to recover a unitary description in situations where we either neglect gravity, or we know how to encode gravitational physics in a unitary theory, for example through AdS/CFT. This is indeed the case in our setup. If we only had the massive worldline, and before coupling to the boundary graviton, we know very well that this integral over times is interpreted as an integral over Schwinger proper time and through LSZ formalism connected to unitary evolution. This continues holding even if we add the observer: as we discussed in section \ref{sec:renormalization}, the propagator of the observer can be mapped to a sum of contribution coming from different unitary QFTs with different masses, see equation \eqref{eq:qftsum}. When we introduce the coupling to the boundary mode however, we understand that there will be effects that are not unitary in the standard sense, since we know that the dynamics of the boundary graviton is mapped in the CFT side to an ensemble of unitary theories, instead of only one. The story would be different if we were in a normal AdS/CFT setup, where the dual is a single boundary theory: the observer's evolution will be encoded in a fully unitary theory on the boundary; while the description in terms of the evolution on the worldline might still be non-unitary.

In this discussion we passed over an important feature of the analytically continued formula \eqref{eq:overlaptauintegral}.  The fact that each operator for fixed $\tau$ is unitary relies on the measure being real for real $\tau$. This is not true in our case, and can be seen easily in the saddle point; the analytic continuation of the saddle points derived in \ref{sec:saddlepoint} are complex, so the overlap contains an exponentially decaying factor. This renders non-unitary the operator that governs the evolution of the observer, even in the semiclassical regime. However, this is an expected feature of our setup; since there is no timelike geodesic that connects two points on the same boundary, the probability to throw in some state from some point on the boundary and measure it from some other point on the same boundary correspond to a quantum process with exponentially suppressed probability.
In particular, in the weak backreaction regime and at large $t$, the amplitude scales in the following way
\begin{align}
	\mel{\psi_f}{e^{-i \tau^* H}}{\psi_i}\approx \mel{\psi_f}{e^{-2 \log\!\left(
\frac{\beta}
{2\pi \epsilon}
\right)H} e^{-i 2 \pi H} e^{-\frac{2\pi t}{\beta}H}}{\psi_i};
\end{align}
Rewriting this in the energy basis and using $\Delta_n\simeq m+E_n$ at large $m$, one finds
\begin{align}
	\braket{\psi_f(t_2)}{\psi_i(t_1)}& \propto e^{-i \tau^* m}\mel{\psi_f}{e^{-i \tau^* H}}{\psi_i} = \\
	&=\sum_n \psi_f^*(E_n)\psi_i(E_n) \epsilon^{2(m+E_n)} e^{\left(-2 \log\!\left(
\frac{\beta}
{2\pi}
\right)-i 2 \pi \right) (m+E_n)} e^{-\frac{2 \pi t}{\beta}(m+E_n)}.
\end{align}
In each term, the cutoff dependent factor has the same form as the scaling of a boundary correlator. The remaining piece has the characteristic decay of semiclassical two point functions of matter fields: each term falls off as $e^{-\frac{2\pi t}{\beta}(m+E_n)}$. Thus, the unnormalized overlap in the probe regime decays exponentially with $t$, in a way that depends on the observer's state. In contrast, the exact gravitationally dressed propagator we derived allows us to study the late-time behavior away from the semiclassical limit. Either from the analytically continued propagator or from the known form of dressed Schwarzian correlators, one finds that the amplitude scales as
\begin{align}
	\braket{\psi_f(t_2)}{\psi_i(t_1)}\propto \frac{\beta^{3/2}}{t^{3/2}(\beta-it)^{3/2}}\sum_n \epsilon^{2\Delta_n}\psi_f^*(E_n)\psi_i(E_n).
\end{align}
This replacement of the exponentially decay of the semiclassical case with this power law decay is analogous to what happens for gravitationally dressed CFT correlators \cite{Altland:2016cdf, Yang:2018gdb, Mertens:2017mtv}. The power $t^{-3}$ is universal and does not depend on the details of the observer's state, while the coefficient still does. It would be interesting to understand how this behavior changes after resumming higher genus contributions, for example by performing the calculation directly in the matrix model.

\subsection{Correlation functions on the worldline} \label{sec:correlators}
In the previous sections we computed the Euclidean propagator of the observer. Other interesting observables are correlation functions of the quantum mechanical degree of freedom on the worldline. We will show in this section that, in our Euclidean setup, the only effect of coupling to gravity is that it is not possible to define uniquely the state in which correlation functions are evaluated. This is the same effect that we observed in the Euclidean propagator. The reason is that, when computing a given correlator on the worldline (for example, a two point function), it is possible to fix the proper time between the operator insertions in a gauge invariant way. What remains of gravity is that it makes the total evolution before and after the insertion fluctuate, inducing the effect described above. Before moving to the derivation, let us make some comments. When defining correlation functions of fields on a worldline coupled to dynamical gravity, it is necessary to use a clock degree of freedom to fix time reparameterizations along the worldline, as was pointed out in \cite{Witten:2023background}. This ensures that there is no diffeomorphism that shifts the worldline time by $\tau \to \tau+\text{const}$. In our setup, however, we are anchoring the worldline at the boundary, so we can fix this reparameterization by fixing proper time to be zero when the quantum system starts evolving from the first boundary point\footnote{Notice that, more generally, one can still define correlation functions between two or more points on the worldline in a gauge invariant way just by fixing the proper time between the insertions. When doing so, of course one must consider diffeos that shift proper time by a constant, and integrate over those.}. This allows us to define correlation functions using the following proper time coordinate:
 \begin{align}
	\sigma(\bar \tau)= \int_0^{\bar \tau} e_\tau d\tau\,.
\end{align}
where of course $\sigma \in (0,\betaobs)$. With the units in \eqref{eq:worldlineunits}, this coordinate already has dimensions of length. Of course, one can add additional degrees of freedom to the worldline that can be used as a clock (as done in \cite{Witten:2023background}) and replace this proper time $\sigma$ with a dynamical clock variable.

If we now change coordinates from $\tau$ to $\sigma$ defined as above, the path integral that defines the propagator $\mel{\psi_f}{U_{QG}}{\psi_i}$ becomes
\begin{align}
	&\bra{\psi_f}U_{QG}\ket{\psi_i}=\int dq_i dq_f \psi_i(q_i)\psi_f^*(q_f) \,\int_0^\infty d\betaobs \mu(\betaobs) \times \\
	&\times \int_{q_i}^{q_f} \mathcal D q \exp\Big\{- \int_0^{\betaobs} d \sigma \, \left(\frac{ \Dot{q}^2}{2}+ V(q)\right)\Big\}
\end{align}
where $\betaobs$ is the total proper time, $\betaobs= \int_0^1 e_\tau d\tau$. To define correlation functions of operators on the worldline, say at proper times $\sigma_1 < \sigma_2 < \ldots < \sigma_n$, we should condition that the system has evolved for at least $\sigma_n$ Euclidean time. One way of defining such conditional correlation functions is then
\begin{align}
	&\bra{\psi_f}q(\sigma_n)\dots q(\sigma_1)\ket{\psi_i}_{QG}=\mathcal{N}_{\sigma_n}\int dq_i dq_f \psi_i(q_i)\psi_f^*(q_f) \,\int_{\sigma_n}^\infty d\betaobs \mu(\betaobs) \times \\
	&\times \int_{q_i}^{q_f} \mathcal D q  \,q(\sigma_n)\dots q(\sigma_1)\,\exp\Big\{- \int_0^{\betaobs} d \sigma \, \left(\frac{ \Dot{q}^2}{2}+ V(q)\right)\Big\},
\end{align}
where
\begin{align}
    \mathcal{N}_{\sigma_n}^{-1}=\int_{\sigma_n}^\infty d\betaobs \mu(\betaobs)\bra{\psi_f}e^{-\betaobs H_{QM}}\ket{\psi_i}\,.
\end{align}
In the Hamiltonian language this corresponds to computing
\begin{align}
	&\bra{\psi_f}q(\sigma_n)\dots q(\sigma_1)\ket{\psi_i}_{QG}=\\
    &=\frac{1}{\int_{\sigma_n}^\infty d\betaobs \mu(\betaobs)\bra{\psi_f}e^{-\betaobs H}\ket{\psi_i}}\int_{\sigma_n}^\infty d\betaobs \mu(\betaobs) \bra{\psi_f}e^{-\betaobs H}  \hat q(\sigma_n)\dots \hat q(\sigma_1)  \ket{\psi_i}\,.
\end{align}
where $\hat q(\sigma)=e^{\sigma H_{QM}}\hat q e^{- \sigma H_{QM}}$ is the standard Heisenberg picture operator. Given an initial state, the coupling to quantum gravity makes the Euclidean time difference between the last insertion and the final state fluctuating: equivalently, it maps a single final states to a superposition of states evolved from $\ket{\psi_f}$ with an operator $e^{-\betaobs H_{QM}}$. This is precisely the same effect that we observed in the Euclidean propagator, and nothing more. If $\mu(\betaobs)$ is peaked at some very large value of $\betaobs$, the correlation function will be dominated by the contribution from the final state being the vacuum and the coupling to quantum gravity does not produce any effect. We can avoid this by working in the regime $\omega \log (1/\epsilon) \ll 1$, as we discussed in section \ref{sec:renormalization}, which is roughly the same regime for which $\omega \betaobs^*$ is small and the quantum system is sensitive to these quantum effects.
\section{Observer's quantum gravity partition function on the double trumpet}
So far we have mostly considered situations where the worldline of the observer is an open segment connecting two points in the asymptotic boundary and the observables are overlaps of states or correlation functions\footnote{With the exception of section \ref{sec:pickup}, which discussion we expand here.}. Moreover, we have seen that the effect of quantum gravity is suppressed by logarithmic functions of the holographic cutoff. If we want to ask questions about the thermodynamics of the observer in the presence of quantum gravity, it is natural to ask that the observer worldline closes inside Euclidean space, so that we are computing a partition function from the point of view of the observer. Another advantage of considering such a setup is that we do not have to worry about holographic renormalization, our answer has a clear quantum gravitational interpretation already as it is. Motivated by this, we discuss a different class of topologies that have been studied in JT gravity, namely the double trumpet \cite{Saad:2018bqo}. One can think of this as one contribution (of many) to a UV regulated\footnote{Since we exclude the disk with closed paths, where one would obtain divergencies from very small Euclidean trajectories.} partition function of the observer in Euclidean JT gravity with arbitrary boundaries, where the worldlines are closed curves. Moreover, this is one of the few topologies in JT gravity where it is possible to find semiclassical solutions, either by adding couplings of the two boundaries, by using complex fields, or by changing boundary conditions \cite{Maldacena:2018lmt, Saad:2018bqo, Garcia-Garcia:2020ttf, Chen:2020tes,  Fumagalli:2024msi}.

We consider the setup depicted in figure~\ref{fig:double-trumpet-observer}, where the worldline is closed and wraps the neck of the double trumpet. As in the previous subsections, we rescale $\beta$ to absorb $\phi_r$. For simplicity of presentation we will also restrict to the case where the two boundaries have equal lengths $\beta$. It is possible to integrate out the wiggly boundary curves and obtain an effective action for the double-trumpet neck length $b$. We stress that this is different from the total worldline proper time $\betaobs$. The path integral is given by
\begin{figure}
	\centering
	\includegraphics[width=0.65\linewidth]{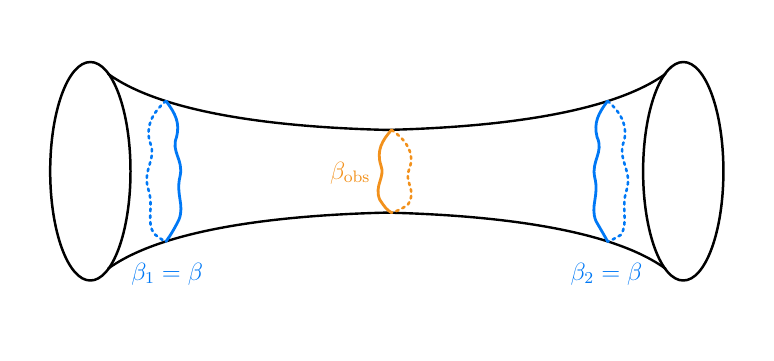}
	\caption{Double-trumpet setup with equal boundaries $\beta_1=\beta_2=\beta$ and a closed worldline wrapping the neck. The orange loop denotes the proper time cycle $\beta_{\rm obs}$ along the lab/worldline, which represents the inverse temperature seen by the observer. The size of the neck is denoted by $b$ and it is in general different from the lab/worldline's proper time.}
	\label{fig:double-trumpet-observer}
\end{figure}
\begin{align}
	Z_{QG}[\beta,m]&=\int_0^\infty d b\, b \int \mathcal Dx_b^{(2)}  e^{-I_{JT}[x_b^{(2)}, b]}\int \mathcal Dx_b^{(1)}  e^{-I_{JT}[x_b^{(1)}, b]}= \\
	&= \frac{1}{\beta}\int_0^\infty d b\, b\,e^{-\frac{b^2}{\beta}}\,.
\end{align}
The measure of the $b$ integral $b\,d b$ comes from the symplectic measure, and in the last step we performed the path integral over the boundary modes \cite{Saad:2018bqo}. Let us now add the quantum system. As in the disk topology (eq. \eqref{eq:propagatorfirst}), the path integral now becomes
\begin{align}
	Z_{QG}[\beta,m]&= \frac{1}{\beta}\int_0^\infty d b\, b \int \mathcal De_\tau \int \mathcal Dx_b^L  e^{-I_{JT}[x_b^L, b]}\int \mathcal Dx_b^R  e^{-I_{JT}[x_b^R, b]}\times \\
	&\times \int_{\text{closed paths}} \mathcal{D}{X_w(\tau)} \int_{q\text{ periodic}} \mathcal D q  e^{-I_{obs}[X_w^\mu(\tau),q(\tau),e_\tau]}\,.
\end{align}
where the path integral over $X_w$ is over closed paths on the trumpet geometry whose neck length is $b$. Upon gauge fixing $e_\tau= \betaobs$, so that $\betaobs$ is the proper time around the closed worldline, and therefore the inverse temperature seen by the observer, we get\footnote{
The einbein gauge fixing for a closed worldline has a residual translation zero mode and gives the measure $d \betaobs/\betaobs$.}
\begin{align}
	Z_{QG}[\beta,m]=&\frac{1}{\beta}\int_0^\infty d b\,b \int_0^\infty \frac{d\betaobs}{\betaobs}\,e^{-\frac{b^2}{\beta}}\nonumber\\
	&\times \int \mathcal DX_w \exp{-m  \int_0^1 d \tau  \left( \frac{\dot X_w^2}{2\betaobs}+ \frac{\betaobs}{2}\right)} Z_{QM}(\betaobs)\,,
\end{align}
where $Z_{QM}(\betaobs)$ is the standard partition function of the quantum system at inverse temperature $\betaobs$. Let us first work in a limit where the $X_w$ path integral localizes to a geodesic, namely the one wrapping the neck; this requires taking $m \gg 1$ (in units of $r_{\rm{AdS}}$). If we do so, the path integral over $X_w$ localizes and the on shell action contributes to the exponent as $\int d\tau \Dot X_w^2/\betaobs= b^2/\betaobs$. There will be fluctuations around the saddle point similar to the $y$ fluctuations in the disk case. This can be taken into account by looking at the heat kernel of AdS$_2$ with a quotient, which describes the double trumpet geometry. Unlike in the disk case, here we can scale the mass to be arbitrarily large without running into problems with backreaction and finite cutoff effects, so that these fluctuations are very suppressed. We therefore remain with
\begin{align}
	Z_{QG}[\beta,m]=\frac{1}{\beta}\int_0^\infty d b\,b \int_0^\infty \frac{d\betaobs}{\betaobs}\,e^{-\frac{b^2}{\beta}}  e^{-\frac{m \betaobs}{2}-\frac{m \, b^2}{2 \betaobs}} Z_{QM}(\betaobs)\,.
\end{align}
This integral does not have a saddle point, as expected, since backreaction from normal matter does not make such geometries on shell \cite{Maldacena:2018lmt}. However, we can consider these geometries as off shell contributions and integrate over all of them to get the effective partition function of the observer coupled to quantum gravity. We can do the integral over the neck length $b$, getting a formula similar to what we had in the disk case:
\begin{align}\label{eq:trumpetmeasure}
	Z_{QG}[\beta,m]=\frac{1}{\beta}\int_0^\infty d\beta_{\rm obs}\frac{e^{-\frac{m \beta_{\rm obs}}{2}}}{ \frac{2\beta_{\rm obs}}{\beta} + m}Z_{QM}(\beta_{\rm obs})= \int_0^\infty d\beta_{\rm obs} \,\mu_{\rm DT}(\beta_{\rm obs}) Z_{QM}(\beta_{\rm obs})\,,
\end{align}
where we use $\mu_{\rm DT}$ to distinguish the measure from the previous one. Notice that after having integrated out $b$, the terms in the exponent that were responsible for the Brownian scaling are gone, so we can interpret the fluctuations as coming directly from the fluctuations of the gravitational mode, which in this case is the size of the neck.

The formula \eqref{eq:trumpetmeasure} allows us to discuss our proposal on how to compute thermodynamic quantities when we take into account the coupling to quantum gravity. Our calculation shows that this coupling induces an integral over inverse temperatures felt by the observer $\beta_{\rm obs}$. As we have anticipated in section \ref{sec:pickup}, one proposal is to first form the usual canonical quantities by differentiating with respect to $\beta_{\rm obs}$, and then average over the quantum gravity measure
\begin{align}
	\expval{\mathcal O}_{QG}\equiv \frac{1}{\mathcal N_{\rm DT}}\int_0^\infty d\beta_{\rm obs}\,\mu_{\rm DT}(\beta_{\rm obs})\,\mathcal O(\beta_{\rm obs})\,,
	\qquad
	\mathcal N_{\rm DT}\equiv \int_0^\infty d\beta_{\rm obs}\,\mu_{\rm DT}(\beta_{\rm obs})\,.
\end{align}
where $\mathcal O(\beta_{\rm obs})=\expval{\mathcal O(\beta_{\rm obs})}_{th}$ is the standard thermal average of a generic thermodynamic quantity $\mathcal O$. For example, the free energy and heat capacity dressed by quantum gravity are
\begin{align}
	\expval{F(\beta_{\rm obs})}_{QG}
	=-\expval{\frac{1}{\beta_{\rm obs}}\log Z_{QM}(\beta_{\rm obs})}_{QG}\,,\\
	\expval{C(\beta_{\rm obs})}_{QG}
	=\expval{\beta_{\rm obs}^2\partial_{\beta_{\rm obs}}^2\log Z_{QM}(\beta_{\rm obs})}_{QG}\,.
\end{align}
Equivalently, for any smooth thermodynamic quantity $\mathcal O(\beta_{\rm obs})$, the correction around the mean value of the induced inverse temperature is controlled by the moments of the same measure,
\begin{align}
	\expval{\mathcal O}_{QG}
	=\mathcal O(\expval{\beta_{\rm obs}})+\frac{1}{2}\expval{(\beta_{\rm obs}-\expval{\beta_{\rm obs}})^2}_{QG}\,
	\partial_{\beta_{\rm obs}}^2\mathcal O(\beta_{\rm obs})\Big|_{\beta_{\rm obs}=\expval{\beta_{\rm obs}}}+\dots\,,
\end{align}
where $\expval{\beta_{\rm obs}}$ is the average value of the inverse temperature. If we now assume that the quantum system does not backreact strongly, we can characterize the effects of quantum gravity by looking at the measure $\mu_{\rm DT}(\beta_{\rm obs})$ and computing the average value of $\beta_{\rm obs}$ and variance around it. Let us start from the average
\begin{align}
	\expval{\beta_{\rm obs}}\equiv \expval{\beta_{\rm obs}}_{QG}=\frac{\int_0^\infty d\beta_{\rm obs} \mu_{\rm DT}(\beta_{\rm obs})  \beta_{\rm obs}}{\int_0^\infty d\beta_{\rm obs} \mu_{\rm DT}(\beta_{\rm obs}) }=\frac{-\frac{2 e^{-\frac{m^2\beta}{4}}}{\text{Ei}\left(-\frac{m^2\beta}{4}\right)}-\frac{m^2\beta}{2}}{m},
\end{align}
with Ei defined as
\begin{align}
	Ei(z)=-\int_{-z}^{\infty } \frac{e^{-t}}{t} \, dt.
\end{align}
We can then compute the average of the fluctuations
\begin{align}
	\expval{(\beta_{\rm obs}-\expval{\beta_{\rm obs}})^2}_{QG}=\frac{e^{-\frac{m^2\beta}{2}} \left(-4-e^{\frac{m^2\beta}{4}} \left(4+m^2\beta\right) \text{Ei}\left(-\frac{m^2\beta}{4}\right)\right)}{m^2 \text{Ei}\left(-\frac{m^2\beta}{4}\right)^2}.
\end{align}
Notice that these expressions are controlled by the quantity
\begin{align}
\lambda_{\rm DT}\equiv m^2\beta\,,
\end{align}
which is different from the disk case, where the corresponding parameter was $m\beta$. This is not surprising, since the coupling to the gravitational mode is different in the two cases. Taking this parameter large or small corresponds, respectively, to strong or weak backreaction of the worldline on the gravitational mode the worldline is coupled to, which in this case is the size of the neck. Let us start from the weak backreaction limit $\lambda_{\rm DT} \ll 1$:
\begin{align}
	&\expval{\beta_{\rm obs}}=\frac{2}{m \log \left(\frac{1}{\lambda_{\rm DT} }\right)}-\frac{2 (\log (4)-\gamma_E )}{m \log ^2(\frac{1}{\lambda_{\rm DT}} )}\\
	&\expval{(\beta_{\rm obs}-\expval{\beta_{\rm obs}})^2}_{QG}=\frac{4}{m^2 \log \left(\frac{1}{\lambda_{\rm DT} }\right)}+\frac{4 (\gamma_E -1-\log (4))}{m^2 \log ^2(\frac{1}{\lambda_{\rm DT}} )}\\
	&\frac{\expval{(\beta_{\rm obs}-\expval{\beta_{\rm obs}})^2}_{QG}}{\expval{\beta_{\rm obs}}^2}=\log{\frac{1}{\lambda_{\rm DT}}}+\text{const}
\end{align}
Already in this weak backreaction regime, the relative fluctuations are parametrically large: the variance is enhanced with respect to $\expval{\beta_{\rm obs}}^2$ by a logarithm. This is unlike the disk case, where the measure has a smooth semiclassical saddle and the fluctuations of the worldline's effective temperature are suppressed. The opposite, strong backreaction limit gives
\begin{align}
	&\expval{\beta_{\rm obs}}=\frac{2}{m}-\frac{8}{\lambda_{\rm DT}  m}\\
	&\expval{(\beta_{\rm obs}-\expval{\beta_{\rm obs}})^2}_{QG}=\frac{4}{m^2}-\frac{32}{\lambda_{\rm DT}  m^2}\\
	&\frac{\expval{(\beta_{\rm obs}-\expval{\beta_{\rm obs}})^2}_{QG}}{\expval{\beta_{\rm obs}}^2}=1+\frac{32}{\lambda_{\rm DT} ^2}
\end{align}
which has order one relative fluctuations. Thus, the double-trumpet observer never enters a regime in which $\beta_{\rm obs}$ is sharply peaked: in the weak-backreaction expansion the variance is parametrically larger than $\expval{\beta_{\rm obs}}^2$, while in the strong-backreaction expansion it remains of the same order. This occurs because the path integral has no smooth semiclassical saddle controlling $\beta_{\rm obs}$\footnote{One could try to find a saddle by going to the microcanonical ensemble, as in \cite{Saad:2018bqo}, but one would find that the saddle is the double cone, which is a singular geometry, with $\beta_{\rm obs}=0$.}, but provides a setup where the fluctuations felt by the observer are not parametrically suppressed, and the results are independent of the cutoff of JT gravity.

\raggedbottom

\section{Conclusions}
In this work we laid the groundwork for studying the effects of quantum gravity on an observer living on a worldline in the bulk of a two dimensional spacetime with dynamical gravity. In particular, we studied the coupling of the observer to the boundary graviton in Jackiw-Teitelboim gravity, at fixed topology. We found that this coupling induces fluctuations in the Euclidean proper length of the observer's worldline, or equivalently in the inverse temperature seen by the observer. As a result, evolution with a fixed operator is replaced by an average over such operators weighted by a measure over Euclidean times. While this is familiar in the worldline formulation of QFT and can be used to compute unitary S-matrix observables, the situation here is further complicated by the presence of boundary gravitational modes. 

On the disk topology, and in the semiclassical limit, this measure becomes peaked around a particular Euclidean time, determined by the equations of motion, thus recovering the standard quantum mechanical evolution for the observer. We also analyzed the size of the fluctuations around this semiclassical point, which we take as a measure of how strongly quantum gravity affects the observer's quantum mechanics. These fluctuations have a twofold origin: they come both from fluctuations of the worldline in the bulk, familiar from quantum field theory and present even without gravity, and from genuine gravitational fluctuations. In the disk topology, the QFT fluctuations dominate over the gravitational ones, and they are still suppressed by a logarithmic function of the holographic cutoff. Given the recent work on finite cutoff JT gravity \cite{Iliesiu:2020zld, Stanford:2020qhm, Chaudhuri:2024yau, Griguolo:2021wgy, Griguolo:2025kpi,Griguolo:2026eoi}, it would be interesting to see whether allowing the boundary to fluctuate inside the bulk of spacetime could enhance the gravitational fluctuations for an observer on the disk topology. Moreover, higher topologies could provide another source of enhanced fluctuations, as suggested by their effect on correlation functions \cite{Saad:2019pqd, Iliesiu:2024cnh}. 

Even when these fluctuations remain suppressed, however, we discussed how they can be detected by an observer with a finely spaced density of states. This is also the same regime in which gravitational fluctuations would affect the quantum system if used as a clock to define gauge-invariant observables as in \cite{Witten:2023background,Chandrasekaran:2022cip, Witten:2021unn, Mirbabayi:2023vgl, Kolchmeyer:2024fly, Yang:2025lme}. It would be interesting to explore further the connection with these works, since in our setup we have control over the interaction of the observer with gravity.

The present treatment also has a few limitations. A central one is that we worked mostly in Euclidean signature. It would be interesting to explore more deeply the Lorentzian physics that can be extracted from our calculations, and to ask what they imply for the experience of an observer interacting with a black hole. In particular, how does the observer perceive the horizon? Does it encounter firewalls, or other interesting features of spacetime? Answering these questions will require a better Lorentzian understanding of our calculation, including the normalization of the states whose overlaps we compute, as well as improving our description to allow the observer to cross the horizon.

Since the disk result gives fluctuations that are suppressed by a function of the IR cutoff, we also presented a calculation of the quantum gravity partition function in a situation where the gravitational fluctuations remain large, namely the double trumpet. In this case the measure for the observer's inverse temperature is not sharply peaked, reflecting the absence of a smooth semiclassical saddle controlling the double trumpet path integral. It would be interesting to see whether a better controlled saddle, or a qualitatively different distribution for $\beta_{\rm obs}$, emerges after deforming the problem by coupling to imaginary sources for additional matter fields, as in \cite{Garcia-Garcia:2020ttf, Held:2026bbo}, or by turning on a (possibly complex) chemical potential for the neck size of the double trumpet.

In this exploratory work, we started from a two-dimensional toy model. Nevertheless, our discussion has identified possible handles for approaching similar questions in higher dimensions. One concrete way to make progress comes from leveraging the worldline formulation of QFT, which has been used in recent work to couple matter to quantum gravity in AdS$_3$ (and dS$_3$) \cite{Castro:2023bvo, Castro:2023dxp, Bourne:2025azc}. Another direction comes from the observation that the bulk experience of the observer can be encoded by a tower of operators in the dual CFT. Understanding how this description is embedded in a nearly CFT$_1$ with defects could provide a systematic way to study the observer's bulk experience in higher dimensions through the dual theory and known CFT techniques \cite{Billo:2016cpy}.

Given our cosmological motivation, one of the most interesting future directions is to apply similar methods to understand the experience of an observer in de Sitter space. In particular, the averages over the observer's inverse temperature that appeared in our analysis resonate with some aspects discussed in \cite{Maldacena:2024spf, Chen:2025jqm}, and in our setup we have control over gravitational effects. A concrete toy setting where the methods we developed may apply is provided by centaur geometries \cite{Anninos:2017hhn}, Euclidean geometries that interpolate between a hyperbolic trumpet and a de Sitter patch.

In all of these cases, the presence of global gravitational modes (related to boundary gravitons on the disk topology and the size of the neck in the double-trumpet) induce a form of averaging over unitary bulk evolution that questions the unitary experience of the bulk observer. Understanding this feature in detail for realistic models of cosmological observers, such as ourselves, would be of great interest in the field of quantum cosmology.

\subsection*{Acknowledgments}
It is a pleasure to acknowledge fruitful discussions with Andreas Blommaert, Jan de Boer, Damian Galante, Jildou Hollander, David Kolchmeyer, Diego Liska, Mehrdad Mirbabayi, Boris Post, Erik Verlinde, and Stathis Vitouladitis. AF thanks ICTP Trieste, King's College London, and Queen Mary University for hospitality. AF thanks the participants of the workshop "Observers, wormholes and complex saddles in cosmology" held at EPFL Lausanne for useful discussions.
\appendix
\section{Gauge fixing the einbein}\label{app:einbeinggaugefixing}

In this section we prove that gauge fixing the einbein in the observer action \eqref{eq:observeraction} leaves no additional factors behind. This derivation applies for an open interval, which is relevant for the disk topology, we discuss the difference with closed intervals at the end of the appendix. The part of the path integral that depends on the einbein is
\begin{align}
	 Z_e= \int \frac{\mathcal D e  }{\text{diff}} \exp{-m\int_0^1 d \tau \,e_\tau\left(e_\tau^{-2}\, \frac{\Dot{X}_w^2}{2}+\frac{1}{2}\right)- \int_0^1 d \tau \,e_\tau\left(e_\tau^{-2}\, \frac{ \Dot{q}^2}{2}+ \frac{\omega^2 {q}^2}{2}\right)}.
\end{align}
In general, this action is invariant under time reparameterization:
\begin{align}
	&\tau \to \sigma (\tau)\\
	& e_\tau \to e_\tau (\partial_\tau \sigma)^{-1}\\
	& \partial_\tau \to (\partial_\tau \sigma)^{-1} \partial_\tau \,.
\end{align}
We take the time reparametrizations to preserve the interval: $\sigma(0)=0\,,\sigma(1)=1$. The gauge invariant measure for the fluctuations of the einbein is
\begin{align}
    \norm{\delta e}^2=\int d\tau\, e_\tau^{-1} (\delta e_\tau)^2\,,
\end{align}
where $\delta e_\tau$ is a functional variation of the einbein. We can perform a change of time coordinate to land on a constant value of the einbein $e_\tau \to e_\tau (\partial_\tau \sigma)^{-1}=\betaobs$ for each value of the gauge invariant modulus $\int d \tau e_\tau=\betaobs$. The generic einbein is therefore related to this gauge fixed one as $e_\tau=\betaobs\,\partial_\tau \sigma$. The functional variation of the generic einbein is therefore related to the variation of $\betaobs$ and $\sigma$ as
\begin{align}
	\delta e_\tau= \delta(\betaobs \partial_\tau \sigma)=\delta \betaobs \partial_\tau \sigma+ \betaobs \partial_\tau \delta\sigma\,.
\end{align}
From this we get the invariant norm
\begin{align}
	\norm{\delta e_\tau}^2
	&=\frac{(\delta \betaobs)^2}{\betaobs}
	+ \int \frac{d\tau}{\betaobs\,\partial_\tau \sigma}\,\betaobs^2 (\partial_\tau \delta \sigma)^2
	+ 2\betaobs\,\delta \betaobs \int \frac{d\tau}{\partial_\tau \sigma}\,
	\partial_\tau \sigma\, \partial_\tau (\delta \sigma) \\
	&=\frac{(\delta \betaobs)^2}{\betaobs}
	+ \int d\tau\, \betaobs (\partial_\tau \delta \sigma)^2
	 \, .
\end{align}
where in the last step we used the fact that we are computing the fluctuations around $\sigma=\tau$ so that $\partial_\tau \sigma=1$, and the last term vanishes since it becomes a boundary term where $\sigma$ is kept fixed. We now expand in modes. The generic decomposition of $\delta \sigma$ is
\begin{align}
	\delta \sigma =\sum_{n>0} \delta \sigma_n \sin(n \pi \tau)\,,
\end{align}
since we keep $\delta \sigma=0$ at the boundaries. We get
\begin{align}
    \norm{\delta e_\tau}^2&=\frac{(\delta \betaobs)^2}{\betaobs}+ \sum_{n \neq 0} \betaobs (2 \pi n)^2 \delta \sigma_n^2 \,.
\end{align}
and therefore the path integral measure for the einbein becomes
\begin{align}
    \mathcal{D}e= \frac{d\betaobs}{\sqrt{\betaobs}} \prod_{n >0} \sqrt{\betaobs} (2 \pi n) d\sigma_n= \frac{d\betaobs}{\betaobs^{3/4}}\prod_{n>0}d \sigma_n\,,
\end{align}
where we used zeta function regularization and dropped numerical factors. We have to divide this gauge fixed measure by the volume of the diffeos, which is just
\begin{align}
	\int \mathcal D\sigma\,.
\end{align}
To do this we need the measure $\mathcal D\sigma$, which again we can get from the invariant norm in the space of functions for $\sigma$:
\begin{align}
	\norm{\delta \sigma}^2= \int d \tau e_\tau^3 (\delta \sigma)^2= \sum_{n>0} \betaobs^3(\delta \sigma_n)^2\,,
\end{align}
which implies
\begin{align}
	 \mathcal{D} \sigma =\prod_{n>0} \betaobs^{3/2} d \sigma_n=\frac{1}{\betaobs^{3/4}}\prod_{n>0} d \sigma_n\,.
\end{align}
We now see that in the ratio all the factors of $\betaobs$ cancel
\begin{align}
	\int \frac{\mathcal D e  }{\text{diff}} = \int d\betaobs \,.
\end{align}
Thus, with the convention used in the main text, the gauge slice is simply $e_\tau=\betaobs$ and the modulus $\betaobs$ is the observer's worldline proper Euclidean time. For a closed worldline, the gauge fixing is similar, but there is a residual translation zero mode that gives the measure $d \betaobs/\betaobs$ instead of $d\betaobs$.

\section{Different actions for the quantum system} \label{app:differentactions}
It turns out that there are different ways to couple the observer to the worldline and the metric in the bulk of the spacetime in which they live. The reason is that there are various possible gauge invariant actions that one can write. In this section we take the observer to be described by a simple harmonic oscillator for concreteness; one can consider a more generic action and the discussion would be unaffected. Let us first consider the perhaps simplest way (which is not the one we considered in the paper) to write a gauge invariant worldline action that describes both the worldline and the quantum system that lives on it:
\begin{align}\label{eq:ho_action_generic}
    I_{obs,1}=m_0\int d\tau \sqrt{g_{\tau \tau}}+\int d \tau \sqrt{g_{\tau \tau}} \left[ \frac{\Dot{q}^2}{2g_{\tau\tau} }+\frac{ \omega^2 q^2}{2}\right],
\end{align}
where $g_{\tau \tau}=g_{\mu \nu}\dot x^\mu \dot x^\nu$. As in the text, we take the coordinate $\tau$ to be dimensionless, so the units are the ones explained in eq. \eqref{eq:worldlineunits}: $g_{\tau \tau}$ has dimensions $(\text{length})^{2}$, and $\omega$ is the characteristic energy scale with dimensions $(\text{length})^{-1}$. With our normalization of $q$, $\omega$ is the only new scale introduced by the quantum system. The action we considered in the main text is instead defined through the einbein $e_\tau$ and reads
\begin{align}\label{eq:ho_action_einbein}
    I_{obs,2}=m\int d \tau \,e_\tau\left(e_\tau^{-2}\, \frac{{\Dot X_w}^2}{2}+\frac{1}{2}\right)+ \int d \tau \,e_\tau\left(e_\tau^{-2}\, \frac{ \Dot{q}^2}{2}+ \frac{\omega^2 {q}^2}{2}\right).
\end{align}
Notice that a shift of the potential energy $\delta E$ of the quantum system in the action $I_{obs,1}$ can be reabsorbed by a redefinition of the mass $m_0\to m_0+\delta E$. In the action $I_{obs,2}$ an analogous shift produces $m^2 \to m^2+2 m \delta E$, leading to the relation of the conformal dimension with the mass and the energy we observed in section \ref{sec:renormalization}. Let us see if this equation is related to the other one \textit{classically}. We can solve the constraint equation for $e_\tau$, getting
\begin{align}
	e_\tau=\left(\frac{m\dot X_w^2+\dot q ^2}{m+\omega^2 q^2}\right)^{1/2}\,.
\end{align}
and
\begin{align}
	 I_{obs,2}=\int d \tau \,\left[(m{\Dot X_w}^2+\Dot{q}^2)(m+\omega^2 q^2)\right]^{1/2}\,.
\end{align}
If the kinetic energy of the harmonic oscillator is much smaller than the worldline coordinate energy, and its potential energy is much smaller than the worldline mass scale,
\begin{align}
	 \Dot{q}^2 \ll m\Dot{X_w}^2 \,,\quad \omega^2 q^2 \ll m
\end{align}
we can expand the action to get ($g_{\tau \tau}\equiv\dot X^2$)
\begin{align}
	I_{obs,2}= m \int d\tau \sqrt{g_{\tau \tau}}+ \int d\tau  \sqrt{g_{\tau \tau}} \left[\frac{\dot q^2}{2g_{\tau \tau}}+\frac{\omega^2 q^2}{2}\right]\,.
\end{align}
which is exactly $I_{obs,1}$ if we identify the masses $m=m_0$.

This equality holds at the classical level and with weak coupling of the quantum system. At the quantum level, the two actions are not equivalent, since they have different path integral measures, one could expect that they are related by a relation involving some UV cutoff, similar to the one discussed in chapter 9 of Polyakov's book \cite{Polyakov:1987ez}, but dependent on the parameters of the observer. The action we considered in the main text is more convenient for our purposes, since it allows us to separate the worldline and quantum system contributions, and to integrate out the worldline in a simple way.

\section{Matching saddle points with solutions of the equations of motion}\label{app:EOM}
In this appendix, we show that the solutions of the saddle point equations of the Euclidean propagator analyzed in section \ref{sec:saddlepoint} match the solutions of the equations of motion.

If one considers the disk, solutions of pure JT gravity are circles at constant values of $r$ in polar coordinates
\begin{align}\label{eq:poincarediskmetric}
	ds^2=dr^2+\sinh(r)^2 d\varphi^2\,,\quad \varphi \in (0,2 \pi)\,.
\end{align}
Solutions of this type are just labeled by the radius of the circle $r=r_c$, which is in turn fixed by the boundary length and the boundary value of the dilaton,
\begin{align}
	\beta_b=2 \pi \sinh(r_c)\,,\quad \phi_b=A \cosh r_c\,.
\end{align}
Using $\beta_b=\beta/\epsilon$ and $\phi_b=\phi_r/\epsilon$, and working with the same temperature rescaled by $\phi_r$ as in the main text, we set $\phi_r=1$ in this appendix. In these conventions, the boundary conditions become $\beta/\epsilon=2 \pi \sinh(r_c)$ and $1/\epsilon=A \cosh r_c$.

We should consider that AdS$_2$ has an SL$(2,\mathbb R)$ isometry group that is inherited by the Schwarzian theory. This means that we have a full SL$(2,\mathbb R)$ worth of solutions, obtained by boosting the above one. For our case, we must consider the more general setup where we add the worldline. As in the main text, we will stick to the case where the quantum system does not backreact, so that we can approximate the worldline with a geodesic of a particle of mass $m$. The worldline backreacts on the dilaton in the bulk, inducing a jump in its derivative normal to the geodesic:
\begin{align}
	\partial_n \phi_2-\partial_n \phi_1=m\,,
\end{align}
where the subscript $1$ labels the right side of the geodesic and the subscript $2$ labels the left side, matching the convention of the main text. The direction of the normal derivative does not matter. The value of the dilaton remains continuous:
\begin{align}
	\phi_2- \phi_1=0\,.
\end{align}
We can construct a cutout AdS$_2$ solution by smoothly gluing two separate cutouts of two hyperbolic disks. We label the right cutout by $1$ and the left cutout by $2$. These cutouts are further cut along a geodesic with some particular induced metric $g_{\text{ind},1},g_{\text{ind},2}$. If we glue these two cutouts by identifying the geodesic and imposing
\begin{align}
	g_{\text{ind},1}=g_{\text{ind},2}\,,
\end{align}
and the dilaton equation, we have constructed a solution. We are therefore left with the problem of matching two Poincaré disks of the form \eqref{eq:poincarediskmetric}, with radii $r_1,r_2$, and with a cut along a geodesic with a particular induced metric on the geodesic; see figure~\ref{fig:poincare-disk-matching}. At the boundaries, the conditions are the usual ones:
\begin{figure}
    \centering
    \includegraphics[width=0.72\linewidth]{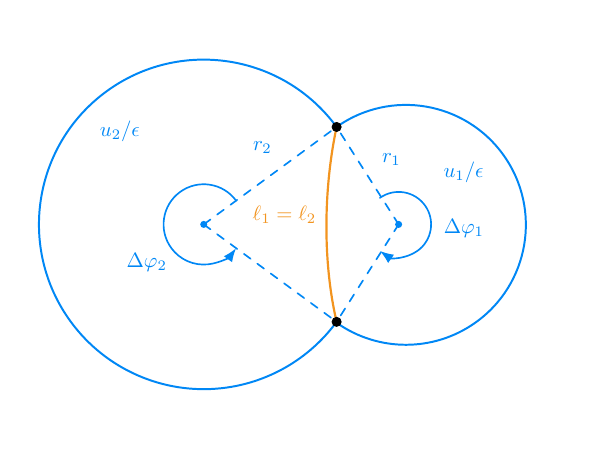}
    \caption{Matching two Poincaré disk cutouts across a common geodesic. The right cutout is labeled $1$ and the left cutout is labeled $2$. The boundary arcs have cutoff lengths $u_1/\epsilon$ and $u_2/\epsilon$, the disk radii are $r_1,r_2$, the angular openings are $\Delta\varphi_1,\Delta\varphi_2$, and the induced metric matching fixes the common geodesic length $\ell_1=\ell_2$.}
    \label{fig:poincare-disk-matching}
\end{figure}
\begin{align}\label{eq:boundaryconditionspolar}
	&A_1\cosh r_1=A_2\cosh r_2=\frac{1}{\epsilon}\,,\\
	&u_1/\epsilon= a(r_1) \Delta \varphi_1\,,\\
	&u_2/\epsilon= a(r_2) \Delta \varphi_2\,.
\end{align}
where $\Delta \varphi$ is the angular extension of the circle. For the matching conditions along the worldline, we instead use another coordinate system:
\begin{align}\label{eq:rhochicoordinates}
	ds^2= d\rho^2+\cosh (\rho)^2 \,d\chi^2\,, \quad \phi= A \cosh(\rho) \cosh(\chi).
\end{align}
This is obtained from the embedding coordinates
\begin{align}
	ds^2= -dT^2+ dX^2+dY^2, \quad \phi= A\, T\,,\quad -T^2+ X^2+Y^2=-1\,, T>0,
\end{align}
with
\begin{align}
	T= \cosh \rho \cosh \chi \,,\,X= \cosh \rho \sinh \chi\,,\,Y= \sinh \rho.
\end{align}
The previous coordinates were obtained instead from
\begin{align}
	T= \cosh r\,,\,X= \sinh r \cos \varphi\,,\,Y= \sinh r \sin \varphi
\end{align}
which one can use to relate the two coordinate systems. In the coordinates \eqref{eq:rhochicoordinates}, slices of constant $\chi$ are geodesics whose closest point from the origin is at distance $\chi=\bar \chi$, and $\rho$ runs along such a geodesic. In these coordinates, the induced metric condition implies that $\rho_1=\rho_2$. Then the conditions for the dilaton, for some constant values of $\chi_1$ and $\chi_2$, are
\begin{align}
	&A_1 \cosh(\chi_1)=A_2 \cosh(\chi_2)\,,\\
	&A_1 \sinh(\chi_1)+A_2 \sinh(\chi_2)=m
\end{align}
These must be added to the equations for the boundary conditions \eqref{eq:boundaryconditionspolar}, which we wrote using polar coordinates. In such coordinates, geodesics satisfy $\cos \varphi_i= \tanh(\chi_i)/\tanh r_i$. If we set $\varphi_i=0$ at the point of minimum approach, the angular distance is $\Delta \varphi_i=2 \pi- 2 \varphi_i$, where
 \begin{align}
 	\varphi_i=\arccos(\frac{\tanh \chi_i}{\tanh r_i}) \simeq \arccos(\tanh \chi_i)\,,\quad i=1,2\,,
 \end{align}
where in the last line we used $r_i\gg 1$. These boundary conditions become
  \begin{align}
 	&u_1/\epsilon = \left(2 \pi-2\arccos(\tanh \chi_1)\right)a(r_1)\,,\\
 	&u_2/\epsilon = \left(2 \pi-2\arccos(\tanh \chi_2)\right)a(r_2)\,.
 \end{align}
We can now substitute $a(r_i)= \sinh r_i \simeq \frac{1}{A_i \epsilon}$, for $i=1,2$, to obtain the final set of equations that we will solve:
\begin{align}\label{eq:EOMmassivechi}
	&u_1= \left(2 \pi-2\arccos(\tanh \chi_1)\right) \frac{1}{A_1 }\,,\\
	&u_2= \left(2 \pi-2\arccos(\tanh \chi_2)\right) \frac{1}{A_2 }\,,\\
	&A_1 \cosh(\chi_1)=A_2 \cosh(\chi_2)\,,\\
	&A_1 \sinh(\chi_1)+A_2 \sinh(\chi_2)=m\,.
\end{align}
These are supplemented by the two equations on the dilaton
 \begin{align}
 	\frac{1}{\epsilon}=A_1 \cosh r_1=A_2 \cosh r_2
 \end{align}
that will fix $r_{1,2}$. The equations \eqref{eq:EOMmassivechi} can also be expressed in terms of the length of the geodesic $d$ or a parameter $\ell$ defined as $\cosh d\simeq \frac{\ell^2}{2 \epsilon^2}$ (for small $\epsilon$), by means of the hyperbolic law of cosines:
 \begin{align}
 	\cosh r_i=\cosh(\frac{d_i}{2})\cosh(\chi_i)\,,\quad i=1,2\,.
 \end{align}
which for large values of $d(\ell), r_i$ becomes
  \begin{align}
 	\cosh r_i=\frac{\ell_i}{2 \epsilon}\cosh(\chi_i)\, \implies \cosh \chi_i =\frac{2}{ \ell_i A_i}\,,\quad i=1,2\,.
 \end{align}
Using this, we reduce the above equations to
\begin{align}\label{eq:EOMmassiveell}
	&u_1= \left(2 \pi-2\arcsin(\frac{A_1 \ell_1}{2})\right) \frac{1}{A_1}\,,\\
	&u_2= \left(2 \pi-2\arcsin(\frac{A_2 \ell_2}{2})\right) \frac{1}{A_2}\,,\\
	&\ell_1=\ell_2\equiv \ell,,\\
	&A_1 \sqrt{\frac{4}{A_1^2 \ell^2}-1}+A_2 \sqrt{\frac{4}{A_2^2 \ell^2}-1}=m\,.
\end{align}
We can solve these equations in various limits. For the large backreaction limit $m u_i\gg 1$,
\begin{align}
	\ell=\frac{4}{m}-\frac{16 \pi^2}{m} \left( \frac{1}{( m u_1)^2} + \frac{1}{( m u_2)^2} \right)+\order{\frac{1}{m}\left((m u_1)^{-4}+(m u_2)^{-4}\right)}
\end{align}
which is the same result obtained in the same limit by solving the saddle point equations \eqref{eq:solutionslargembeta}. For small backreaction $m u_i\ll 1$ we have to proceed more carefully, since we can solve the equations only perturbatively around the symmetric case. In the exactly symmetric case $u_1=u_2=\beta/2$ and small $m u_i$, we get
\begin{align}
 	\ell=\frac{\beta }{\pi }-\frac{\beta ^2 m}{2 \pi ^3}
\end{align}
which again matches the result of the main text around \eqref{eq:solutionssmallmbeta_l}. Expanding around the symmetric solution for $A_1,A_2$, etc., we get ($\beta=u_1+u_2$)

\begin{equation}
	\ell=\frac{\beta}{\pi}+\frac{\pi (\beta - 2 u_1)^2}{8 \beta}
	+ \left( -\frac{\beta^2}{2 \pi^3} + \frac{\pi}{32} (\beta - 2 u_1)^2 \right) m
\end{equation}

This, expanded around $u_1=\beta/2+x$, gives
\begin{align}
	\ell=\frac{\beta}{\pi}+x^2 \left(\frac{\pi }{2 \beta }+\frac{\pi  m}{8}\right)-\frac{\beta ^2 m}{2 \pi ^3}
\end{align}
which also matches the solution \eqref{eq:solutionssmallmbeta_l} if expanded around the symmetric solution.
\endgroup
\bibliography{main}

@article{Freivogel:2026ofo,
    author = "Freivogel, Ben and Speranza, Antony and Verlinde, Erik",
    title = "{Quantum Fluctuations of the Black Hole Horizon}",
    eprint = "2606.28243",
    archivePrefix = "arXiv",
    primaryClass = "hep-th",
    month = "6",
    year = "2026"
}

@article{Freivogel:2026bsx,
    author = "Freivogel, Ben and Moitra, Upamanyu",
    title = "{Large Quantum Gravity Fluctuations of BTZ Black Holes}",
    eprint = "2606.28160",
    archivePrefix = "arXiv",
    primaryClass = "hep-th",
    month = "6",
    year = "2026"
}

@article{Verlinde:2022hhs,
    author = "Verlinde, Erik and Zurek, Kathryn M.",
    title = "{Modular fluctuations from shockwave geometries}",
    eprint = "2208.01059",
    archivePrefix = "arXiv",
    primaryClass = "hep-th",
    doi = "10.1103/PhysRevD.106.106011",
    journal = "Phys. Rev. D",
    volume = "106",
    number = "10",
    pages = "106011",
    year = "2022"
}

@article{Verlinde:2019ade,
    author = "Verlinde, Erik and Zurek, Kathryn M.",
    title = "{Spacetime Fluctuations in AdS/CFT}",
    eprint = "1911.02018",
    archivePrefix = "arXiv",
    primaryClass = "hep-th",
    doi = "10.1007/JHEP04(2020)209",
    journal = "JHEP",
    volume = "04",
    pages = "209",
    year = "2020"
}

@article{Verlinde:2019xfb,
    author = "Verlinde, Erik P. and Zurek, Kathryn M.",
    title = "{Observational signatures of quantum gravity in interferometers}",
    eprint = "1902.08207",
    archivePrefix = "arXiv",
    primaryClass = "gr-qc",
    doi = "10.1016/j.physletb.2021.136663",
    journal = "Phys. Lett. B",
    volume = "822",
    pages = "136663",
    year = "2021"
}

@article{Marolf:2003bb,
    author = "Marolf, Donald",
    editor = "Trampetic, Josip and Wess, Julius",
    title = "{On the quantum width of a black hole horizon}",
    eprint = "hep-th/0312059",
    archivePrefix = "arXiv",
    doi = "10.1007/3-540-26798-0_9",
    journal = "Springer Proc. Phys.",
    volume = "98",
    pages = "99--112",
    year = "2005"
}

@article{Billo:2016cpy,
    author = "Bill{\`o}, Marco and Gon{\c{c}}alves, Vasco and Lauria, Edoardo and Meineri, Marco",
    title = "{Defects in conformal field theory}",
    eprint = "1601.02883",
    archivePrefix = "arXiv",
    primaryClass = "hep-th",
    doi = "10.1007/JHEP04(2016)091",
    journal = "JHEP",
    volume = "04",
    pages = "091",
    year = "2016"
}

@article{Griguolo:2026eoi,
    author = "Griguolo, Luca and Papalini, Jacopo and Russo, Lorenzo and Seminara, Domenico and Tarana, Alex",
    title = {{Quantum JT Gravity in a box as a P{\"o}schl-Teller Scattering Problem}},
    eprint = "2607.01385",
    archivePrefix = "arXiv",
    primaryClass = "hep-th",
    month = "7",
    year = "2026"
}

@article{Franken:2026bff,
    author = "Franken, Victor and Mertens, Thomas G. and de S. L. Torres, Bruno",
    title = "{Falling through the horizon of a quantum black hole}",
    eprint = "2607.03344",
    archivePrefix = "arXiv",
    primaryClass = "hep-th",
    month = "7",
    year = "2026"
}

@article{Banihashemi:2026mje,
    author = "Banihashemi, Batoul and Batra, Gauri and Law, Albert Y. T. and Silverstein, Eva and Torroba, Gonzalo",
    title = "{The yes boundaries wavefunctions of the universe}",
    eprint = "2604.10267",
    archivePrefix = "arXiv",
    primaryClass = "hep-th",
    month = "4",
    year = "2026"
}

@article{Philcox:2025faf,
    author = "Philcox, Oliver H. E. and Silverstein, Eva and Torroba, Gonzalo",
    title = "{Quantum stress-energy at timelike boundaries: Testing a new beyond-{\ensuremath{\Lambda}}CDM parameter with cosmological data}",
    eprint = "2507.00115",
    archivePrefix = "arXiv",
    primaryClass = "astro-ph.CO",
    doi = "10.1103/162y-rmsv",
    journal = "Phys. Rev. D",
    volume = "113",
    number = "4",
    pages = "043548",
    year = "2026"
}

@article{Silverstein:2024xnr,
    author = "Silverstein, Eva and Torroba, Gonzalo",
    title = "{Timelike-bounded dS$_{4}$ holography from a solvable sector of the T$^{2}$ deformation}",
    eprint = "2409.08709",
    archivePrefix = "arXiv",
    primaryClass = "hep-th",
    doi = "10.1007/JHEP03(2025)156",
    journal = "JHEP",
    volume = "03",
    pages = "156",
    year = "2025"
}

@article{Batra:2024kjl,
    author = "Batra, Gauri and De Luca, G. Bruno and Silverstein, Eva and Torroba, Gonzalo and Yang, Sungyeon",
    title = "{Bulk-local dS$_{3}$ holography: the matter with $ T\overline{T} $ + {\ensuremath{\Lambda}}$_{2}$}",
    eprint = "2403.01040",
    archivePrefix = "arXiv",
    primaryClass = "hep-th",
    doi = "10.1007/JHEP10(2024)072",
    journal = "JHEP",
    volume = "10",
    pages = "072",
    year = "2024"
}

@article{Coleman:2021nor,
    author = "Coleman, Evan and Mazenc, Edward A. and Shyam, Vasudev and Silverstein, Eva and Soni, Ronak M. and Torroba, Gonzalo and Yang, Sungyeon",
    title = "{De Sitter microstates from T$ \overline{T} $ + {\ensuremath{\Lambda}}$_{2}$ and the Hawking-Page transition}",
    eprint = "2110.14670",
    archivePrefix = "arXiv",
    primaryClass = "hep-th",
    doi = "10.1007/JHEP07(2022)140",
    journal = "JHEP",
    volume = "07",
    pages = "140",
    year = "2022"
}

@article{Anninos:2024xhcadsbdy,
    author = "Anninos, Dionysios and Arias, Ra{\'u}l and Galante, Dami{\'a}n A. and Maneerat, Chawakorn",
    title = "{Gravitational observatories in AdS$_{4}$}",
    eprint = "2412.16305",
    archivePrefix = "arXiv",
    primaryClass = "hep-th",
    doi = "10.1007/JHEP07(2025)234",
    journal = "JHEP",
    volume = "07",
    pages = "234",
    year = "2025"
}

@article{Anninos:2024wpycosmoobs,
    author = "Anninos, Dionysios and Galante, Dami{\'a}n A. and Maneerat, Chawakorn",
    title = "{Cosmological observatories}",
    eprint = "2402.04305",
    archivePrefix = "arXiv",
    primaryClass = "hep-th",
    doi = "10.1088/1361-6382/ad5824",
    journal = "Class. Quant. Grav.",
    volume = "41",
    number = "16",
    pages = "165009",
    year = "2024"
}

@article{Anninos:2023epigravobs,
    author = "Anninos, Dionysios and Galante, Dami{\'a}n A. and Maneerat, Chawakorn",
    title = "{Gravitational observatories}",
    eprint = "2310.08648",
    archivePrefix = "arXiv",
    primaryClass = "hep-th",
    doi = "10.1007/JHEP12(2023)024",
    journal = "JHEP",
    volume = "12",
    pages = "024",
    year = "2023"
}

@article{Bourne:2025azc,
    author = "Bourne, Robert and Fliss, Jackson R. and Knighton, Bob",
    title = "{A spool for every quotient: One-loop partition functions in AdS$_3$ gravity}",
    eprint = "2507.05364",
    archivePrefix = "arXiv",
    primaryClass = "hep-th",
    doi = "10.21468/SciPostPhys.20.3.065",
    journal = "SciPost Phys.",
    volume = "20",
    pages = "065",
    year = "2026"
}

@article{Castro:2023bvo,
    author = "Castro, Alejandra and Coman, Ioana and Fliss, Jackson R. and Zukowski, Claire",
    title = "{Coupling Fields to 3D Quantum Gravity via Chern-Simons Theory}",
    eprint = "2304.02668",
    archivePrefix = "arXiv",
    primaryClass = "hep-th",
    doi = "10.1103/PhysRevLett.131.171602",
    journal = "Phys. Rev. Lett.",
    volume = "131",
    number = "17",
    pages = "171602",
    year = "2023"
}

@article{Castro:2023dxp,
    author = "Castro, Alejandra and Coman, Ioana and Fliss, Jackson R. and Zukowski, Claire",
    title = "{Keeping matter in the loop in dS$_{3}$ quantum gravity}",
    eprint = "2302.12281",
    archivePrefix = "arXiv",
    primaryClass = "hep-th",
    doi = "10.1007/JHEP07(2023)120",
    journal = "JHEP",
    volume = "07",
    pages = "120",
    year = "2023",
    note = "[Erratum: JHEP 09, 004 (2024)]"
}

@article{Chen:2025jqm,
    author = "Chen, Yiming and Stanford, Douglas and Tang, Haifeng and Yang, Zhenbin",
    title = "{On the phase of the de Sitter density of states}",
    eprint = "2511.01400",
    archivePrefix = "arXiv",
    primaryClass = "hep-th",
    doi = "10.1007/JHEP05(2026)068",
    journal = "JHEP",
    volume = "05",
    pages = "068",
    year = "2026"
}

@article{Maldacena:2024spf,
    author = "Maldacena, Juan",
    title = "{Real observers solving imaginary problems}",
    eprint = "2412.14014",
    archivePrefix = "arXiv",
    primaryClass = "hep-th",
    month = "12",
    year = "2024"
}

@article{Griguolo:2025kpi,
    author = "Griguolo, Luca and Papalini, Jacopo and Russo, Lorenzo and Seminara, Domenico",
    title = "{A new perspective on dilaton gravity at finite cutoff}",
    eprint = "2512.21774",
    archivePrefix = "arXiv",
    primaryClass = "hep-th",
    month = "12",
    year = "2025"
}

@article{Mertens:2017mtv,
    author = "Mertens, Thomas G. and Turiaci, Gustavo J. and Verlinde, Herman L.",
    title = "{Solving the Schwarzian via the Conformal Bootstrap}",
    eprint = "1705.08408",
    archivePrefix = "arXiv",
    primaryClass = "hep-th",
    doi = "10.1007/JHEP08(2017)136",
    journal = "JHEP",
    volume = "08",
    pages = "136",
    year = "2017"
}

@article{Camporesi:1990wm,
    author = "Camporesi, R.",
    title = "{Harmonic analysis and propagators on homogeneous spaces}",
    doi = "10.1016/0370-1573(90)90120-Q",
    journal = "Phys. Rept.",
    volume = "196",
    pages = "1--134",
    year = "1990"
}

@article{Witten:2021unn,
    author = "Witten, Edward",
    title = "{Gravity and the crossed product}",
    eprint = "2112.12828",
    archivePrefix = "arXiv",
    primaryClass = "hep-th",
    doi = "10.1007/JHEP10(2022)008",
    journal = "JHEP",
    volume = "10",
    pages = "008",
    year = "2022"
}

@article{Leutheusser:2021frk,
    author = "Leutheusser, Samuel Aaron Wehlau and Liu, Hong",
    title = "{Emergent Times in Holographic Duality}",
    eprint = "2112.12156",
    archivePrefix = "arXiv",
    primaryClass = "hep-th",
    reportNumber = "MIT-CTP/5382",
    doi = "10.1103/PhysRevD.108.086020",
    journal = "Phys. Rev. D",
    volume = "108",
    number = "8",
    pages = "086020",
    year = "2023"
}

@article{Leutheusser:2021qhd,
    author = "Leutheusser, Samuel and Liu, Hong",
    title = "{Causal connectability between quantum systems and the black hole interior in holographic duality}",
    eprint = "2110.05497",
    archivePrefix = "arXiv",
    primaryClass = "hep-th",
    reportNumber = "MIT-CTP/5335",
    doi = "10.1103/PhysRevD.108.086019",
    journal = "Phys. Rev. D",
    volume = "108",
    number = "8",
    pages = "086019",
    year = "2023"
}

@article{Jafferis:2020ora,
    author = "Jafferis, Daniel Louis and Lamprou, Lampros",
    title = "{Inside the hologram: reconstructing the bulk observer{\textquoteright}s experience}",
    eprint = "2009.04476",
    archivePrefix = "arXiv",
    primaryClass = "hep-th",
    doi = "10.1007/JHEP03(2022)084",
    journal = "JHEP",
    volume = "03",
    pages = "084",
    year = "2022"
}

@article{deBoer:2022zps,
    author = "de Boer, Jan and Jafferis, Daniel Louis and Lamprou, Lampros",
    title = "{On black hole interior reconstruction, singularities and the emergence of time}",
    eprint = "2211.16512",
    archivePrefix = "arXiv",
    primaryClass = "hep-th",
    month = "11",
    year = "2022"
}

@article{Blommaert:2026ofx,
    author = "Blommaert, Andreas and Tietto, Damiano and Verlinde, Herman",
    title = "{An observer's quantization of 3d de Sitter}",
    eprint = "2606.26241",
    archivePrefix = "arXiv",
    primaryClass = "hep-th",
    month = "6",
    year = "2026"
}

@article{Goto:2026ipq,
    author = "Goto, Kanato and Milekhin, Alexey and Verlinde, Herman and Xu, Jiuci",
    title = "{Generalized Free Fields in de Sitter from 1D CFT}",
    eprint = "2605.03037",
    archivePrefix = "arXiv",
    primaryClass = "hep-th",
    reportNumber = "OU-HET-1309, RIKEN-iTHEMS-Report-26",
    month = "5",
    year = "2026"
}

@article{Tietto:2025oxn,
    author = "Tietto, Damiano and Verlinde, Herman",
    title = "{A microscopic model of de Sitter spacetime with an observer}",
    eprint = "2502.03869",
    archivePrefix = "arXiv",
    primaryClass = "hep-th",
    month = "2",
    year = "2025"
}

@article{Verlinde:2024zrh,
    author = "Verlinde, Herman and Zhang, Mengyang",
    title = "{SYK correlators from 2D Liouville-de Sitter gravity}",
    eprint = "2402.02584",
    archivePrefix = "arXiv",
    primaryClass = "hep-th",
    doi = "10.1007/JHEP05(2025)053",
    journal = "JHEP",
    volume = "05",
    pages = "053",
    year = "2025"
}

@article{Verlinde:2024znh,
    author = "Verlinde, Herman",
    title = "{Double-scaled SYK, chords and de Sitter gravity}",
    eprint = "2402.00635",
    archivePrefix = "arXiv",
    primaryClass = "hep-th",
    doi = "10.1007/JHEP03(2025)076",
    journal = "JHEP",
    volume = "03",
    pages = "076",
    year = "2025"
}

@article{Narovlansky:2023lfz,
    author = "Narovlansky, Vladimir and Verlinde, Herman",
    title = "{Double-scaled SYK and de Sitter holography}",
    eprint = "2310.16994",
    archivePrefix = "arXiv",
    primaryClass = "hep-th",
    doi = "10.1007/JHEP05(2025)032",
    journal = "JHEP",
    volume = "05",
    pages = "032",
    year = "2025"
}

@article{Zhao:2026mpl,
    author = "Zhao, Ying",
    title = "{''It from Bit'': The Hartle-Hawking state and quantum mechanics for de Sitter observers}",
    eprint = "2602.05939",
    archivePrefix = "arXiv",
    primaryClass = "hep-th",
    reportNumber = "MIT-CTP/6000",
    month = "2",
    year = "2026"
}

@article{Harlow:2026hky,
    author = "Harlow, Daniel",
    title = "{Observers, $\alpha$-parameters, and the Hartle-Hawking state}",
    eprint = "2602.03835",
    archivePrefix = "arXiv",
    primaryClass = "hep-th",
    reportNumber = "MIT-CTP/5900",
    month = "2",
    year = "2026"
}

@article{Abdalla:2026mxn,
    author = "Abdalla, Ahmed I. and Antonini, Stefano and Bousso, Raphael and Iliesiu, Luca V. and Levine, Adam and Shahbazi-Moghaddam, Arvin",
    title = "{Consistent Evaluation of the No-Boundary Proposal}",
    eprint = "2602.02682",
    archivePrefix = "arXiv",
    primaryClass = "hep-th",
    month = "2",
    year = "2026"
}

@article{Anninos:2025fer,
    author = "Anninos, Dionysios and Hertog, Thomas and Karlsson, Joel",
    title = "{Quantum Liouville cosmology}",
    eprint = "2512.15969",
    archivePrefix = "arXiv",
    primaryClass = "hep-th",
    doi = "10.1088/1475-7516/2026/06/065",
    journal = "JCAP",
    volume = "06",
    pages = "065",
    year = "2026"
}

@article{Yang:2025lme,
    author = "Yang, Zhenbin and Zhang, Yuzhen and Zheng, Wenwen",
    title = "{Comments on the de Sitter Double Cone}",
    eprint = "2505.08647",
    archivePrefix = "arXiv",
    primaryClass = "hep-th",
    reportNumber = "USTC-ICTS/PCFT-25-58",
    month = "5",
    year = "2025"
}

@article{Ivo:2024ill,
    author = "Ivo, Victor and Li, Yue-Zhou and Maldacena, Juan",
    title = "{The no boundary density matrix}",
    eprint = "2409.14218",
    archivePrefix = "arXiv",
    primaryClass = "hep-th",
    month = "9",
    year = "2024"
}

@article{Fumagalli:2024msi,
    author = "Fumagalli, Alessandro and Gorbenko, Victor and Kames-King, Joshua",
    title = "{De Sitter Bra-Ket wormholes}",
    eprint = "2408.08351",
    archivePrefix = "arXiv",
    primaryClass = "hep-th",
    doi = "10.1007/JHEP05(2025)074",
    journal = "JHEP",
    volume = "05",
    pages = "074",
    year = "2025"
}

@article{Anninos:2024iwf,
    author = {Anninos, Dionysios and Baracco, Chiara and M\"uhlmann, Beatrix},
    title = "{Remarks on 2D quantum cosmology}",
    eprint = "2406.15271",
    archivePrefix = "arXiv",
    primaryClass = "hep-th",
    doi = "10.1088/1475-7516/2024/10/031",
    journal = "JCAP",
    volume = "10",
    pages = "031",
    year = "2024"
}

@article{Chaudhuri:2024yau,
    author = "Chaudhuri, Soumyadeep and Ferrari, Frank",
    title = "{Finite cut-off JT and Liouville quantum gravities on the disk at one loop}",
    eprint = "2404.03748",
    archivePrefix = "arXiv",
    primaryClass = "hep-th",
    doi = "10.1007/JHEP01(2025)160",
    journal = "JHEP",
    volume = "01",
    pages = "160",
    year = "2025"
}

@article{Iliesiu:2024cnh,
    author = "Iliesiu, Luca V. and Levine, Adam and Lin, Henry W. and Maxfield, Henry and Mezei, M{\'a}rk",
    title = "{On the non-perturbative bulk Hilbert space of JT gravity}",
    eprint = "2403.08696",
    archivePrefix = "arXiv",
    primaryClass = "hep-th",
    doi = "10.1007/JHEP10(2024)220",
    journal = "JHEP",
    volume = "10",
    pages = "220",
    year = "2024"
}

@article{Mirbabayi:2023vgl,
    author = "Mirbabayi, Mehrdad",
    title = "{An observer\textquoteright{}s measure of de Sitter entropy}",
    eprint = "2311.07724",
    archivePrefix = "arXiv",
    primaryClass = "hep-th",
    doi = "10.1007/JHEP10(2024)077",
    journal = "JHEP",
    volume = "10",
    pages = "077",
    year = "2024"
}

@article{Goel:2020yxl,
    author = "Goel, Akash and Iliesiu, Luca V. and Kruthoff, Jorrit and Yang, Zhenbin",
    title = "{Classifying boundary conditions in JT gravity: from energy-branes to $\alpha$-branes}",
    eprint = "2010.12592",
    archivePrefix = "arXiv",
    primaryClass = "hep-th",
    doi = "10.1007/JHEP04(2021)069",
    journal = "JHEP",
    volume = "04",
    pages = "069",
    year = "2021"
}

@article{Chen:2020tes,
    author = "Chen, Yiming and Gorbenko, Victor and Maldacena, Juan",
    title = "{Bra-ket wormholes in gravitationally prepared states}",
    eprint = "2007.16091",
    archivePrefix = "arXiv",
    primaryClass = "hep-th",
    doi = "10.1007/JHEP02(2021)009",
    journal = "JHEP",
    volume = "02",
    pages = "009",
    year = "2021"
}

@article{Iliesiu:2020zld,
    author = "Iliesiu, Luca V. and Kruthoff, Jorrit and Turiaci, Gustavo J. and Verlinde, Herman",
    title = "{JT gravity at finite cutoff}",
    eprint = "2004.07242",
    archivePrefix = "arXiv",
    primaryClass = "hep-th",
    doi = "10.21468/SciPostPhys.9.2.023",
    journal = "SciPost Phys.",
    volume = "9",
    pages = "023",
    year = "2020"
}

@article{Maldacena:2019cbz,
    author = "Maldacena, Juan and Turiaci, Gustavo J. and Yang, Zhenbin",
    title = "{Two dimensional Nearly de Sitter gravity}",
    eprint = "1904.01911",
    archivePrefix = "arXiv",
    primaryClass = "hep-th",
    doi = "10.1007/JHEP01(2021)139",
    journal = "JHEP",
    volume = "01",
    pages = "139",
    year = "2021"
}

@article{Yang:2018gdb,
    author = "Yang, Zhenbin",
    title = "{The Quantum Gravity Dynamics of Near Extremal Black Holes}",
    eprint = "1809.08647",
    archivePrefix = "arXiv",
    primaryClass = "hep-th",
    doi = "10.1007/JHEP05(2019)205",
    journal = "JHEP",
    volume = "05",
    pages = "205",
    year = "2019"
}

@article{Saad:2018bqo,
    author = "Saad, Phil and Shenker, Stephen H. and Stanford, Douglas",
    title = "{A semiclassical ramp in SYK and in gravity}",
    eprint = "1806.06840",
    archivePrefix = "arXiv",
    primaryClass = "hep-th",
    month = "6",
    year = "2018"
}

@article{Maldacena:2018lmt,
    author = "Maldacena, Juan and Qi, Xiao-Liang",
    title = "{Eternal traversable wormhole}",
    eprint = "1804.00491",
    archivePrefix = "arXiv",
    primaryClass = "hep-th",
    month = "4",
    year = "2018"
}

@article{Engelsoy:2016xyb,
    author = {Engels\"oy, Julius and Mertens, Thomas G. and Verlinde, Herman},
    title = "{An investigation of AdS$_{2}$ backreaction and holography}",
    eprint = "1606.03438",
    archivePrefix = "arXiv",
    primaryClass = "hep-th",
    doi = "10.1007/JHEP07(2016)139",
    journal = "JHEP",
    volume = "07",
    pages = "139",
    year = "2016"
}

@article{Maldacena:2016upp,
    author = "Maldacena, Juan and Stanford, Douglas and Yang, Zhenbin",
    title = "{Conformal symmetry and its breaking in two dimensional Nearly Anti-de-Sitter space}",
    eprint = "1606.01857",
    archivePrefix = "arXiv",
    primaryClass = "hep-th",
    doi = "10.1093/ptep/ptw124",
    journal = "PTEP",
    volume = "2016",
    number = "12",
    pages = "12C104",
    year = "2016"
}

@article{Bergamin:2015vxa,
    author = "Bergamin, R. and Tseytlin, A. A.",
    title = "{Heat kernels on cone of $AdS_2$ and $k$-wound circular Wilson loop in $AdS_5 \times S^5$ superstring}",
    eprint = "1510.06894",
    archivePrefix = "arXiv",
    primaryClass = "hep-th",
    reportNumber = "IMPERIAL-TP-AT-2015-07",
    doi = "10.1088/1751-8113/49/14/14LT01",
    journal = "J. Phys. A",
    volume = "49",
    number = "14",
    pages = "14LT01",
    year = "2016"
}

@article{Almheiri:2014cka,
    author = "Almheiri, Ahmed and Polchinski, Joseph",
    title = "{Models of AdS$_{2}$ backreaction and holography}",
    eprint = "1402.6334",
    archivePrefix = "arXiv",
    primaryClass = "hep-th",
    doi = "10.1007/JHEP11(2015)014",
    journal = "JHEP",
    volume = "11",
    pages = "014",
    year = "2015"
}

@article{Held:2026bbo,
    author = "Held, Jesse and Kaplan, Molly and Marolf, Donald and Wang, Zhencheng",
    title = "{Lorentzian Path Integrals and Jackiw-Teitelboim wormholes with imaginary scalars}",
    eprint = "2601.09932",
    archivePrefix = "arXiv",
    primaryClass = "hep-th",
    month = "1",
    year = "2026"
}

@article{Mertens:2025rpa,
    author = "Mertens, Thomas G. and Tappeiner, Thomas and de S. L. Torres, Bruno",
    title = "{Fiducial observers and the thermal atmosphere in the black hole quantum throat}",
    eprint = "2507.20983",
    archivePrefix = "arXiv",
    primaryClass = "hep-th",
    doi = "10.1007/JHEP04(2026)145",
    journal = "JHEP",
    volume = "04",
    pages = "145",
    year = "2026"
}

@article{Speranza:2025joj,
    author = "Speranza, Antony J.",
    title = "{An intrinsic cosmological observer}",
    eprint = "2504.07630",
    archivePrefix = "arXiv",
    primaryClass = "hep-th",
    doi = "10.1088/1361-6382/ae134c",
    journal = "Class. Quant. Grav.",
    volume = "42",
    number = "21",
    pages = "215023",
    year = "2025"
}

@article{Akers:2025ahe,
    author = "Akers, Chris and Bueller, Gracemarie and DeWolfe, Oliver and Higginbotham, Kenneth and Reinking, Johannes and Rodriguez, Rudolph",
    title = "{On observers in holographic maps}",
    eprint = "2503.09681",
    archivePrefix = "arXiv",
    primaryClass = "hep-th",
    doi = "10.1007/JHEP05(2025)201",
    journal = "JHEP",
    volume = "05",
    pages = "201",
    year = "2025"
}

@article{Abdalla:2025gzn,
    author = "Abdalla, Ahmed I. and Antonini, Stefano and Iliesiu, Luca V. and Levine, Adam",
    title = "{The gravitational path integral from an observer{\textquoteright}s point of view}",
    eprint = "2501.02632",
    archivePrefix = "arXiv",
    primaryClass = "hep-th",
    doi = "10.1007/JHEP05(2025)059",
    journal = "JHEP",
    volume = "05",
    pages = "059",
    year = "2025"
}

@article{Kolchmeyer:2024fly,
    author = "Kolchmeyer, David K. and Liu, Hong",
    title = "{Chaos and the Emergence of the Cosmological Horizon}",
    eprint = "2411.08090",
    archivePrefix = "arXiv",
    primaryClass = "hep-th",
    reportNumber = "MIT-CTP/5805",
    month = "11",
    year = "2024"
}

@article{Banerjee:2024ioz,
    author = "Banerjee, Archi and Kibe, Tanay and Molina, Mart{\'\i}n and Mukhopadhyay, Ayan",
    title = "{Generalized conformal quantum mechanics as an ideal observer in two-dimensional gravity}",
    eprint = "2409.15415",
    archivePrefix = "arXiv",
    primaryClass = "hep-th",
    doi = "10.1103/PhysRevD.111.066011",
    journal = "Phys. Rev. D",
    volume = "111",
    number = "6",
    pages = "066011",
    year = "2025"
}

@article{Nitti:2024iyj,
    author = "Nitti, Francesco and Piazza, Federico and Taskov, Alexander",
    title = "{Relativity of the event: examples in JT gravity and linearized GR}",
    eprint = "2402.01847",
    archivePrefix = "arXiv",
    primaryClass = "hep-th",
    doi = "10.1007/JHEP10(2024)092",
    journal = "JHEP",
    volume = "10",
    pages = "092",
    year = "2024"
}

@article{Witten:2023xze,
    author = "Witten, Edward",
    title = "{A background-independent algebra in quantum gravity}",
    eprint = "2308.03663",
    archivePrefix = "arXiv",
    primaryClass = "hep-th",
    doi = "10.1007/JHEP03(2024)077",
    journal = "JHEP",
    volume = "03",
    pages = "077",
    year = "2024"
}

@article{Jafferis:2022wez,
    author = "Jafferis, Daniel Louis and Kolchmeyer, David K. and Mukhametzhanov, Baur and Sonner, Julian",
    title = "{Jackiw-Teitelboim gravity with matter, generalized eigenstate thermalization hypothesis, and random matrices}",
    eprint = "2209.02131",
    archivePrefix = "arXiv",
    primaryClass = "hep-th",
    doi = "10.1103/PhysRevD.108.066015",
    journal = "Phys. Rev. D",
    volume = "108",
    number = "6",
    pages = "066015",
    year = "2023"
}

@article{Chandrasekaran:2022cip,
    author = "Chandrasekaran, Venkatesa and Longo, Roberto and Penington, Geoff and Witten, Edward",
    title = "{An algebra of observables for de Sitter space}",
    eprint = "2206.10780",
    archivePrefix = "arXiv",
    primaryClass = "hep-th",
    doi = "10.1007/JHEP02(2023)082",
    journal = "JHEP",
    volume = "02",
    pages = "082",
    year = "2023"
}

@article{Anninos:2021ydw,
    author = "Anninos, Dionysios and Hofman, Diego M. and Vitouladitis, Stathis",
    title = "{One-dimensional Quantum Gravity and the Schwarzian theory}",
    eprint = "2112.03793",
    archivePrefix = "arXiv",
    primaryClass = "hep-th",
    doi = "10.1007/JHEP03(2022)121",
    journal = "JHEP",
    volume = "03",
    pages = "121",
    year = "2022"
}

@article{Griguolo:2021wgy,
    author = "Griguolo, Luca and Panerai, Rodolfo and Papalini, Jacopo and Seminara, Domenico",
    title = "{Nonperturbative effects and resurgence in Jackiw-Teitelboim gravity at finite cutoff}",
    eprint = "2106.01375",
    archivePrefix = "arXiv",
    primaryClass = "hep-th",
    reportNumber = "UUITP-25/21",
    doi = "10.1103/PhysRevD.105.046015",
    journal = "Phys. Rev. D",
    volume = "105",
    number = "4",
    pages = "046015",
    year = "2022"
}

@article{Garcia-Garcia:2020ttf,
    author = "Garc{\'\i}a-Garc{\'\i}a, Antonio M. and Godet, Victor",
    title = "{Euclidean wormhole in the Sachdev-Ye-Kitaev model}",
    eprint = "2010.11633",
    archivePrefix = "arXiv",
    primaryClass = "hep-th",
    doi = "10.1103/PhysRevD.103.046014",
    journal = "Phys. Rev. D",
    volume = "103",
    number = "4",
    pages = "046014",
    year = "2021"
}

@article{Stanford:2020qhm,
    author = "Stanford, Douglas and Yang, Zhenbin",
    title = "{Finite-cutoff JT gravity and self-avoiding loops}",
    eprint = "2004.08005",
    archivePrefix = "arXiv",
    primaryClass = "hep-th",
    month = "4",
    year = "2020"
}

@article{Penington:2019kki,
    author = "Penington, Geoff and Shenker, Stephen H. and Stanford, Douglas and Yang, Zhenbin",
    title = "{Replica wormholes and the black hole interior}",
    eprint = "1911.11977",
    archivePrefix = "arXiv",
    primaryClass = "hep-th",
    doi = "10.1007/JHEP03(2022)205",
    journal = "JHEP",
    volume = "03",
    pages = "205",
    year = "2022"
}

@article{Saad:2019pqd,
    author = "Saad, Phil",
    title = "{Late Time Correlation Functions, Baby Universes, and ETH in JT Gravity}",
    eprint = "1910.10311",
    archivePrefix = "arXiv",
    primaryClass = "hep-th",
    month = "10",
    year = "2019"
}

@article{Blommaert:2019hjr,
    author = "Blommaert, Andreas and Mertens, Thomas G. and Verschelde, Henri",
    title = "{Clocks and Rods in Jackiw-Teitelboim Quantum Gravity}",
    eprint = "1902.11194",
    archivePrefix = "arXiv",
    primaryClass = "hep-th",
    doi = "10.1007/JHEP09(2019)060",
    journal = "JHEP",
    volume = "09",
    pages = "060",
    year = "2019"
}

@article{Anninos:2018svg,
    author = "Anninos, Dionysios and Galante, Dami{\'a}n A. and Hofman, Diego M.",
    title = "{De Sitter horizons {\&} holographic liquids}",
    eprint = "1811.08153",
    archivePrefix = "arXiv",
    primaryClass = "hep-th",
    doi = "10.1007/JHEP07(2019)038",
    journal = "JHEP",
    volume = "07",
    pages = "038",
    year = "2019"
}

@article{Anninos:2017hhn,
    author = "Anninos, Dionysios and Hofman, Diego M.",
    title = "{Infrared Realization of dS$_2$ in AdS$_2$}",
    eprint = "1703.04622",
    archivePrefix = "arXiv",
    primaryClass = "hep-th",
    doi = "10.1088/1361-6382/aab143",
    journal = "Class. Quant. Grav.",
    volume = "35",
    number = "8",
    pages = "085003",
    year = "2018"
}

@article{Maldacena:2016hyu,
    author = "Maldacena, Juan and Stanford, Douglas",
    title = "{Remarks on the Sachdev-Ye-Kitaev model}",
    eprint = "1604.07818",
    archivePrefix = "arXiv",
    primaryClass = "hep-th",
    doi = "10.1103/PhysRevD.94.106002",
    journal = "Phys. Rev. D",
    volume = "94",
    number = "10",
    pages = "106002",
    year = "2016"
}

@article{Mathur:2009hf,
    author = "Mathur, Samir D.",
    editor = "Uranga, A. M.",
    title = "{The Information paradox: A Pedagogical introduction}",
    eprint = "0909.1038",
    archivePrefix = "arXiv",
    primaryClass = "hep-th",
    doi = "10.1088/0264-9381/26/22/224001",
    journal = "Class. Quant. Grav.",
    volume = "26",
    pages = "224001",
    year = "2009"
}

@article{Almheiri:2012rt,
    author = "Almheiri, Ahmed and Marolf, Donald and Polchinski, Joseph and Sully, James",
    title = "{Black Holes: Complementarity or Firewalls?}",
    eprint = "1207.3123",
    archivePrefix = "arXiv",
    primaryClass = "hep-th",
    doi = "10.1007/JHEP02(2013)062",
    journal = "JHEP",
    volume = "02",
    pages = "062",
    year = "2013"
}

@article{Anninos:2011af,
    author = "Anninos, Dionysios and Hartnoll, Sean A. and Hofman, Diego M.",
    title = "{Static Patch Solipsism: Conformal Symmetry of the de Sitter Worldline}",
    eprint = "1109.4942",
    archivePrefix = "arXiv",
    primaryClass = "hep-th",
    doi = "10.1088/0264-9381/29/7/075002",
    journal = "Class. Quant. Grav.",
    volume = "29",
    pages = "075002",
    year = "2012"
}

@article{Goheer:2002vf,
    author = "Goheer, Naureen and Kleban, Matthew and Susskind, Leonard",
    title = "{The Trouble with de Sitter space}",
    eprint = "hep-th/0212209",
    archivePrefix = "arXiv",
    doi = "10.1088/1126-6708/2003/07/056",
    journal = "JHEP",
    volume = "07",
    pages = "056",
    year = "2003"
}

@inproceedings{Banks:2003cg,
    author = "Banks, T.",
    title = "{Some thoughts on the quantum theory of de sitter space}",
    booktitle = "{The Davis Meeting on Cosmic Inflation}",
    eprint = "astro-ph/0305037",
    archivePrefix = "arXiv",
    reportNumber = "DAVISINFLATION-2003-BANKS",
    month = "5",
    year = "2003"
}

@article{Parikh:2004wh,
    author = "Parikh, Maulik K. and Verlinde, Erik P.",
    title = "{De Sitter holography with a finite number of states}",
    eprint = "hep-th/0410227",
    archivePrefix = "arXiv",
    reportNumber = "CU-TP-1122, ITFA-2004-49",
    doi = "10.1088/1126-6708/2005/01/054",
    journal = "JHEP",
    volume = "01",
    pages = "054",
    year = "2005"
}

@article{Banks:2006rx,
    author = "Banks, Tom and Fiol, Bartomeu and Morisse, Alexander",
    title = "{Towards a quantum theory of de Sitter space}",
    eprint = "hep-th/0609062",
    archivePrefix = "arXiv",
    reportNumber = "RUNHETC-06-21, SCIPP-06-11, ITFA-2006-31",
    doi = "10.1088/1126-6708/2006/12/004",
    journal = "JHEP",
    volume = "12",
    pages = "004",
    year = "2006"
}

@article{Altland:2016cdf,
    author = "Bagrets, Dmitry and Altland, Alexander and Kamenev, Alex",
    editor = "Unno, Yoshinobu and Ohsugi, Takashi and Hou, Suen and Sadrozinski, Hartmut F. -W. and Lou, Xinchou and Zhu, Hongbo and Ouyang, Qun",
    title = "{Sachdev\textendash{}Ye\textendash{}Kitaev model as Liouville quantum mechanics}",
    eprint = "1607.00694",
    archivePrefix = "arXiv",
    primaryClass = "cond-mat.str-el",
    doi = "10.1016/j.nuclphysb.2016.08.002",
    journal = "Nucl. Phys. B",
    volume = "911",
    pages = "191--205",
    year = "2016"
}

@article{Witten:2023background,
    author = "Witten, Edward",
    title = "{A background-independent algebra in quantum gravity}",
    eprint = "2308.03663",
    archivePrefix = "arXiv",
    primaryClass = "hep-th",
    doi = "10.1007/JHEP03(2024)077",
    journal = "JHEP",
    volume = "03",
    pages = "077",
    year = "2024"
}

@article{saad2019jt,
      title={{JT} gravity as a matrix integral}, 
      author={Phil Saad and Stephen H. Shenker and Douglas Stanford},
      year={2019},
      eprint={1903.11115},
      archivePrefix={arXiv},
      primaryClass={hep-th}
}

@article{Penington:2019npb,
    author = "Penington, Geoffrey",
    title = "{Entanglement Wedge Reconstruction and the Information Paradox}",
    eprint = "1905.08255",
    archivePrefix = "arXiv",
    primaryClass = "hep-th",
    doi = "10.1007/JHEP09(2020)002",
    journal = "JHEP",
    volume = "09",
    pages = "002",
    year = "2020"
}

@article{Almheiri:2019psf,
    author = "Almheiri, Ahmed and Engelhardt, Netta and Marolf, Donald and Maxfield, Henry",
    title = "{The entropy of bulk quantum fields and the entanglement wedge of an evaporating black hole}",
    eprint = "1905.08762",
    archivePrefix = "arXiv",
    primaryClass = "hep-th",
    doi = "10.1007/JHEP12(2019)063",
    journal = "JHEP",
    volume = "12",
    pages = "063",
    year = "2019"
}

@article{Almheiri:2019qdq,
    author = "Almheiri, Ahmed and Hartman, Thomas and Maldacena, Juan and Shaghoulian, Edgar and Tajdini, Amirhossein",
    title = "{Replica Wormholes and the Entropy of Hawking Radiation}",
    eprint = "1911.12333",
    archivePrefix = "arXiv",
    primaryClass = "hep-th",
    doi = "10.1007/JHEP05(2020)013",
    journal = "JHEP",
    volume = "05",
    pages = "013",
    year = "2020"
}

@article{Maldacena:1997re,
    author = "Maldacena, Juan Martin",
    title = "{The Large N limit of superconformal field theories and supergravity}",
    eprint = "hep-th/9711200",
    archivePrefix = "arXiv",
    reportNumber = "HUTP-97-A097, HUTP-98-A097",
    doi = "10.1023/A:1026654312961",
    journal = "Adv. Theor. Math. Phys.",
    volume = "2",
    pages = "231--252",
    year = "1998"
}

@article{Witten:1998qj,
    author = "Witten, Edward",
    title = "{Anti-de Sitter space and holography}",
    eprint = "hep-th/9802150",
    archivePrefix = "arXiv",
    reportNumber = "IASSNS-HEP-98-15",
    doi = "10.4310/ATMP.1998.v2.n2.a2",
    journal = "Adv. Theor. Math. Phys.",
    volume = "2",
    pages = "253--291",
    year = "1998"
}

@article{Jackiw:1984je,
    author = "Jackiw, R.",
    editor = "Baier, R. and Satz, H.",
    title = "{Lower Dimensional Gravity}",
    reportNumber = "MIT-CTP-1203",
    doi = "10.1016/0550-3213(85)90448-1",
    journal = "Nucl. Phys. B",
    volume = "252",
    pages = "343--356",
    year = "1985"
}

@article{Teitelboim:1983ux,
    author = "Teitelboim, C.",
    title = "{Gravitation and Hamiltonian Structure in Two Space-Time Dimensions}",
    doi = "10.1016/0370-2693(83)90012-6",
    journal = "Phys. Lett. B",
    volume = "126",
    pages = "41--45",
    year = "1983"
}

@article{Almheiri:2020cfm,
    author = "Almheiri, Ahmed and Hartman, Thomas and Maldacena, Juan and Shaghoulian, Edgar and Tajdini, Amirhossein",
    title = "{The entropy of Hawking radiation}",
    eprint = "2006.06872",
    archivePrefix = "arXiv",
    primaryClass = "hep-th",
    doi = "10.1103/RevModPhys.93.035002",
    journal = "Rev. Mod. Phys.",
    volume = "93",
    number = "3",
    pages = "035002",
    year = "2021"
}

@article{Hartle:1983ai,
    author = "Hartle, J. B. and Hawking, S. W.",
    editor = "Fang, Li-Zhi and Ruffini, R.",
    title = "{Wave Function of the Universe}",
    reportNumber = "PRINT-83-0937 (CAMBRIDGE)",
    doi = "10.1103/PhysRevD.28.2960",
    journal = "Phys. Rev. D",
    volume = "28",
    pages = "2960--2975",
    year = "1983"
}

@book{Polyakov:1987ez,
    author = "Polyakov, Alexander M.",
    title = "{Gauge Fields and Strings}",
    publisher = "Harwood Academic Publishers",
    address = "Chur, Switzerland",
    series = "Contemporary Concepts in Physics",
    volume = "3",
    pages = "1--301",
    isbn = "978-3-7186-0393-0",
    year = "1987"
}
\bibliographystyle{utphys}
\end{document}